\documentclass[preprint,showpacs,preprintnumbers,amsmath,amssymb,nofootinbib]{revtex4}

\usepackage{graphicx}
\usepackage{dcolumn}
\usepackage{bm}

\begin{document}


\title{Orbifold Family Unification in SO(2N) Gauge Theory}

\author{Yoshiharu Kawamura}
 \email{haru@azusa.shinshu-u.ac.jp}
\author{Takashi Miura}%
 \email{s09t302@shinshu-u.ac.jp}
\affiliation{%
Department of Physics, Shinshu University, Matsumoto, Nagano 390-8621, Japan\\
}%


\date{\today}

\begin{abstract}
We study the possibility of family unification on the basis of $SO(2N)$ gauge theory
on the five-dimensional space-time, $M^4\times S^1/Z_2$.
Several $SO(10)$, $SU(4) \times SU(2)_L \times SU(2)_R$ or $SU(5)$ multiplets
come from a single bulk multiplet of $SO(2N)$ after the orbifold breaking.
Other multiplets including brane fields are necessary to compose
three families of quarks and leptons.
\end{abstract}


\maketitle

\section{Introduction}

The {\it family unification} or {\it flavor unification} based on a large symmetry group 
can provide a possible solution for the origin of the family replication~\cite{R,G,F,K&Y}.
However, we encounter difficulty in the unification on the four-dimensional Minkowski space,
because of extra fields such as $\lq\lq$mirror particles" existing in the higher-dimensional representation.
The mirror particles are particles with opposite quantum numbers under the standard model (SM) gauge group.
If the idea of family or flavor unification is to be realized in nature, extra particles must disappear 
from the low-energy spectrum around the weak scale.
Several interesting mechanisms have been proposed to get rid of the unwelcomed particles.
One is to adopt the $\lq\lq$survival hypothesis", which is the assumption that 
{\it if a symmetry is broken down to a smaller symmetry at a scale $M_{SB}$,
then any fermion mass terms invariant under the smaller group induce fermion masses of order $O(M_{SB})$}~\cite{G,BNM&S}.
Georgi investigated whether an anomaly free set of no-repeated representations in $SU(N)$ models 
can lead to families based on the survival hypothesis, 
and found that three families are derived from $[11, 4] + [11, 8] + [11, 9] + [11, 10]$ in $SU(11)$ model in four dimensions~\cite{G}.
Another possibility is to confine extra particles at a high-energy scale by some strong interaction~\cite{GR&S}.

If we move from four dimensions to higher dimensions, 
there is a possibility to reduce substances including mirror particles using the symmetry reduction 
concerning extra dimensions, as originally discussed in superstring theory~\cite{CHS&W,DHV&W}.
Hence it is meaningful to re-examine the idea of family or flavor unification using grand unified theories (GUTs)
on a higer-dimensional space-time.\footnote{
Five-dimensional supersymmetric GUTs on $M^4 \times S^1/Z_2$ possess the attractive feature that
the triplet-doublet splitting of Higgs multiplets is elegantly realized~\cite{K,H&N}.}
We refer to the family unification using orbifolds for extra dimensions as the orbifold family unification.
There are several preceding studies on the orbifold family unification.
The complete family unification has been suggested in $E_8$ GUT on $M^4 \times T^2/Z_3$~\cite{BB&K}.
The model that three families come from a combination of a bulk gauge multiplet and a few brane fields
in $SO(10)$ GUT on $M^4 \times T^2/Z_3$ has been examined~\cite{Watari:2002tf}.
The gauge, Higgs and matter unification has been proposed in $SU(8)$ GUT on $M^4 \times T^2/Z_6$~\cite{GMN}
and $M^4 \times T^2/Z_3$~\cite{GLMS} and $SO(16)$ on $M^4 \times T^2/Z_6$~\cite{GMN2}.
The orbifold family unification has been studied in $SU(N)$ on $M^4 \times S^1/Z_2$~\cite{KK&O}.

In this paper, we study the possibility of orbifold family unification on the basis of $SO(2N)$ gauge theory
on $M^4\times S^1/Z_2$ using the method in Ref.~\cite{KK&O}.\footnote{
$SO(10)$ GUTs on $M^4 \times T^2/Z_2$~\cite{SO10-T2} and $M^4 \times S^1/Z_2$~\cite{SO10-S1} and
$SO(12)$ GUT on $M^4 \times S^1/Z_2$~\cite{SO12-T2} have been constructed 
and their phenomenological implications have been studied.}
We investigate whether or not three families are derived from a single bulk multiplet of $SO(2N)$
for several orbifold symmetry breaking patterns.

The contents of this paper are as follows.
In section II, we review and provide general arguments on the orbifold breaking on $S^1/Z_2$.
In section III, we investigate unification of quarks and leptons in $SO(2N)$ gauge theory on $M^4 \times S^1/Z_2$. 
Section IV is devoted to conclusions and discussions.
We discuss the gauge equivalence of boundary conditions (BCs) in the appendix A
and the symmetry breaking of $SO(2N+1)$ in the appendix B.

\section{$S^1/Z_2$ Orbifold Breaking}

In this section, we study the orbifold symmetry breaking mechanism 
in $SO(2N)$ gauge theory on $M^4 \times S^1/Z_2$, where $M^4$ is the four-dimensional Minkowski space.

\subsection{Boundary conditions and symmetry reduction on $S^1/Z_2$}

First we review the symmetry reduction mechanism on $S^1/Z_2$ briefly~\cite{Z2}.
Let $x$ (or $x^{\mu}$, $\mu = 0, \cdots, 3$) and $y$ (or $x^5$) be coordinates of $M^4$ and $S^1/Z_2$, respectively.
The $S^1/Z_2$ is obtained by dividing the circle $S^1$ (with the identification $y \sim y + 2\pi R$)
by the $Z_2$ transformation $y \to -y$ so that the point $y$ is identified with $-y$.
Here, $R$ is the radius of $S^1$.
Then the $S^1/Z_2$ is regarded as an interval with length $\pi R$.
Both end points $y=0$ and $\pi R$ are fixed points under the $Z_2$ transformation.
For the operations:
\begin{align}
s_0:y \rightarrow -y~,~~
s_1:y \rightarrow 2\pi R-y~,~~
t:y \rightarrow y+2\pi R~,
\label{st}
\end{align} 
the following relations hold:
\begin{align}
s_0^2=s_1^2=I~,~~t=s_1 s_0~,
\label{st-rel}
\end{align}
where $I$ is the identity operation.
The operation $s_1$ is the reflection at the end point $y=\pi R$ and the $S^1/Z_2$ can be defined using $s_0$ and $s_1$.

Although the point $y$ is identified with the points $-y$ and $2\pi R-y$ on $S^1/Z_2$,
a field does not necessarily take an identical value at these points.
We require that the Lagrangian density should be single valued.
Then the following BCs of the field $\Phi(x,y)$ are allowed: 
\begin{align}
& \Phi (x,-y)=T_{\Phi}[s_0]\Phi (x,y)~,~~
\Phi (x,2\pi R-y)=T_{\Phi}[s_1]\Phi (x,y)~, \notag \\
& \Phi (x,y+2\pi R)=T_{\Phi}[t]\Phi (x,y)~,
\label{PhiBC}
\end{align}
where $T_{\Phi}[s_0]$, $T_{\Phi}[s_1]$ and $T_{\Phi}[t]$ represent appropriate representation matrices 
for $s_0$, $s_1$ and $t$ operations, respectively.
The $T_{\Phi}[*]$ belong to the group elements of transformations which keep the action integral invariant
and satisfy the counterparts of (\ref{st-rel}):
\begin{align}
T_{\Phi}[s_0]^2=T_{\Phi}[s_1]^2=I~,~~ T_{\Phi}[t]=T_{\Phi}[s_0]T_{\Phi}[s_1]~,
\label{PhiBC-rel}
\end{align}
where $I$ stands for the unit matrix.
For the eigenstates of $T_{\Phi}[s_0]$ and $T_{\Phi}[s_1]$, 
the eigenvalues are interpreted as the $Z_2$ parity for the fifth coordinate flip and take $+1$ or $-1$ by definition
Then the eigenvalues of $T_{\Phi}[t]$ also take $+1$ or $-1$.
As the assignment of $Z_2$ parity determines BCs of each multiplet on $S^1/Z_2$,
we use $\lq\lq Z_2$ parity" as a parallel expression of $\lq\lq$BCs on $S^1/Z_2$" in the remainder of the paper.

Let $\phi^{(\mathcal{P}_0, \mathcal{P}_1; \mathcal{U})}(x,y)$ be a component in a multiplet $\Phi(x,y)$
and have definite eigenvalues $(\mathcal{P}_0, \mathcal{P}_1; \mathcal{U})$ for $s_0$, $s_1$ and $t$ operations.
The Fourier expansion of $\phi^{(\mathcal{P}_0, \mathcal{P}_1; \mathcal{U})}(x,y)$ is given by
\begin{align}
&\phi^{(++;+)}(x,y)={1\over \sqrt{\pi R}}\phi_0(x)+\sqrt{2\over \pi R}\sum_{n=1}^{\infty}\phi_n(x)\cos{ny\over R}~,
\label{+++} \\
&\phi^{(--;+)}(x,y)=\sqrt{2\over \pi R}\sum_{n=1}^{\infty}\phi_n(x)\sin{ny\over R}~,
\label{--+} \\
&\phi^{(+-;-)}(x,y)=\sqrt{2\over \pi R}\sum_{n=1}^{\infty}\phi_n(x)\cos{(n-{1\over 2})y\over R}~,
\label{+--} \\
&\phi^{(-+;-)}(x,y)=\sqrt{2\over \pi R}\sum_{n=1}^{\infty}\phi_n(x)\sin{(n-{1\over 2})y\over R}~,
\label{-+-}
\end{align}
where $\pm$ indicates the eigenvalues $\pm 1$.

In the above expansions (\ref{+++}) - (\ref{-+-}), the coefficients $\phi_0(x)$ and $\phi_n(x)$ $(n=1,2,\cdots)$ are four-dimensional fields,
 which are called zero mode and Kaluza-Klein (KK) modes, respectively.
The KK modes $\phi_n(x)$ acquire the mass $n/R$ for $(\mathcal{P}_0, \mathcal{P}_1; \mathcal{U})=(\pm1,\pm1;+1)$,
and $(n-\frac{1}{2})/R$ for $(\mathcal{P}_0, \mathcal{P}_1; \mathcal{U})=(\pm1,\mp1;-1)$ upon compactification.
{\it Unless all components of the non-singlet field have a common $Z_2$ parity, 
a symmetry reduction occurs upon compactification because $\phi_0(x)$ are absent in fields with an odd parity.}
This kind of symmetry breaking is called $\lq\lq$orbifold breaking"~\cite{K0}.

Our four-dimensional world is assumed to be a Minkowski space at one of the fixed points, 
on the basis of the $\lq\lq$brane world scenario".
There exist two kinds of four-dimensional fields in our low-energy theory.
One is the brane field which lives only at the boundary, and the other is the zero mode stemming from the bulk field.
The massive modes $\phi_n(x)$ do not appear in our low-energy world 
because they have heavy masses of $O(1/R)$, with the same magnitude as the unification scale.
Chiral anomalies may arise at the boundaries with the advent of chiral fermions.
Those anomalies must be cancelled in the four-dimensional effective theory by the contribution of brane chiral fermions
and/or counterterms such as the Chern-Simons term~\cite{C&H,KK&L}.

\subsection{Orbifold symmetry breaking of $SO(2N)$}


The $SO(2N)$ is the orthogonal group whose determinant is one and number of elements are $N(2N-1)$.
The representation matrices of $SO(2N)$ are expressed as $e^{i\theta^{\alpha} T^{\alpha}}$
where $\theta^{\alpha}$ is a real parameter and $T^{\alpha}$ are elements of the Lie algebra $so(2N)$. 
The generators $T^{\alpha}$ $(\alpha = 1, \cdots N(2N-1))$ are pure imaginary antisymmetric matrices, i.e., 
$(T^{\alpha})^t = -T^{\alpha}$ and $(T^{\alpha})^* = -T^{\alpha}$.
The generators for vector representation ${\bf{2N}}$ are written by the direct product 
of $2 \times 2$ matrix and $N \times N$ matrix:
\begin{align}
&\sigma_2 \otimes S_N~,~\left(\frac{N(N+1)}{2}-1\right)~;~~ \sigma_0 \otimes A_N~,~\left(\frac{N(N-1)}{2}\right)~;~~ 
\sigma_2 \otimes I_N~,~(1)~;
\notag \\
&\sigma_1 \otimes A_N~,~\left(\frac{N(N-1)}{2}\right)~;~~ \sigma_3 \otimes A_N~,~\left(\frac{N(N-1)}{2}\right)~,
\label{so(2N)}
\end{align}
where $\sigma_i$ $(i=1,2,3)$ are Pauli matrices, $\sigma_0$ is the $2 \times 2$ unit matrix,
$S_N$, $A_N$ and $I_N$ stand for $N \times N$ symmetric matrices (whose components are real),
$N \times N$ antisymmetric matrices (whose components are pure imaginary) and the $N \times N$ unit matrix
and the numbers in the parenthesis represent the numbers of elements.
The elements of subalgebra $su(N)$ are $\sigma_2 \otimes S_N$ and $\sigma_0 \otimes A_N$.

As a warming-up, we consider the breakdown of $SO(2N)$ by the $Z_2$ projection 
with the following type of $2N \times 2N$ matrix:
\begin{align}
P =\sigma_0 \otimes I_{m,n}~~~{\mbox{or}}~~~\sigma_2 \otimes I_{m,n}~,
\label{P}
\end{align}
where $I_{m,n}$ is defined by
\begin{align}
I_{m,n} \equiv {\mbox{diag}}(\underbrace{+1, \cdots, +1}_m, \underbrace{-1, \cdots, -1}_{n(=N-m)})~.
\label{Imn}
\end{align}

~~\\
(1) $P =\sigma_0 \otimes I_{m,n}$\\
The generators for unbroken symmetry commute with $P$, i.e., $[P, T^a] = 0$, and they are given by
\begin{align}
&\sigma_2 \otimes S_m~;~~ \sigma_0 \otimes A_m~;~~ 
\sigma_2 \otimes I_m~;~~
\sigma_1 \otimes A_m~;~~ \sigma_3 \otimes A_m~;
\notag \\
&\sigma_2 \otimes S_n~;~~ \sigma_0 \otimes A_n~;~~ 
\sigma_2 \otimes I_n~;~~
\sigma_1 \otimes A_n~;~~ \sigma_3 \otimes A_n~,
\label{so(2m)so(2n)}
\end{align}
where $S_m$ $(S_n)$, $A_m$ $(A_n)$ and $I_m$ $(I_n)$
stand for $m \times m$ $(n \times n)$ symmetric submatrices,
$m \times m$ $(n \times n)$ antisymmetric submatrices and the $m \times m$ $(n \times n)$ unit submatrix. 
Hence the unbroken symmetry is $SO(2m) \times SO(2n)$.

~~\\
(2) $P =\sigma_2 \otimes I_{m,n}$\\
The generators which commute with $P$ are given by
\begin{align}
&\sigma_2 \otimes S_m~;~~ \sigma_0 \otimes A_m~;~~ 
\sigma_2 \otimes I_m~;~~
\sigma_2 \otimes S_n~;~~ \sigma_0 \otimes A_n~;~~ 
\sigma_2 \otimes I_n~;~~
\notag \\
&\sigma_1 \otimes A_{m,n}~;~~ \sigma_3 \otimes A_{m,n}~,
\label{su(N)u(1)}
\end{align}
where $A_{m,n}$ are antisymmetric matrices composed by off-diagonal $m \times n$ and $n \times m$ submatrices
and commute with $I_{m,n}$.
Hence the unbroken symmetry is $SU(N) \times U(1)$.
~~\\

We study the combination of $Z_2$ projections with $\sigma_0 \otimes I_{m,n}$ and $\sigma_2 \otimes I_{m,n}$.
The generators which simultaneously commute with $\sigma_0 \otimes I_{m,n}$ and $\sigma_2 \otimes I_{m,n}$ are given by
\begin{align}
\sigma_2 \otimes S_m~;~~ \sigma_0 \otimes A_m~;~~ 
\sigma_2 \otimes I_m~;~~
\sigma_2 \otimes S_n~;~~ \sigma_0 \otimes A_n~;~~ 
\sigma_2 \otimes I_n~.
\label{su(m)su(n)}
\end{align}
The unbroken symmetry is $SU(m) \times SU(n) \times U(1)^2$.
The same intersections can be obtained with the combination of $\sigma_0 \otimes I_{m,n}$ and $\sigma_2 \otimes I_N$
or that of $\sigma_2 \otimes I_{m,n}$ and $\sigma_2 \otimes I_N$.


We study the BCs in $SO(2N)$ gauge theory on $M^4 \times S^1/Z_2$.
The BCs on $S^1/Z_2$ are specified by the $2N \times 2N$ matrices $(P_0, P_1, U)$ 
where $P_0^2 = P_1^2 = I$ and $U = P_0 P_1$.
For $(P_0, P_1, U)$, we use the following type of matrices:
\begin{align}
P^{(0)} \equiv \sigma_0 \otimes \tilde{I}~~~~\mbox{or}~~~~P^{(2)} \equiv \sigma_2 \otimes \tilde{I}'~,
\label{sp}
\end{align}
where $\tilde{I}$ and $\tilde{I}'$ are $N \times N$ diagonal matrices whose diagonal components take $+1$ or $-1$.
In this case, the relations $P_0 P_1 = P_1 P_0 = U$ and $U^2 = I$ hold
and the symmetry breaking patterns are classified into following two types.

~~\\
(Type-I)~~All matrices belong to $P^{(0)}$ type.
By the arrangment of the rows and columns, $(P_0, P_1, U)$ are written by
\begin{align}
(P_0 ,P_1 ,U)
&=(\sigma_0 \otimes \tilde{I}_1, \sigma_0 \otimes \tilde{I}_2, \sigma_0 \otimes \tilde{I}_3)~,
\label{typeI}
\end{align}
where $\tilde{I}_1$, $\tilde{I}_2$ and $\tilde{I}_3(=\tilde{I}_1 \tilde{I}_2)$ are defined by
\begin{align}
& \tilde{I}_1 \equiv{\mbox{diag}}(\overbrace{+1, \cdots,+1,+1, \cdots, +1, -1, \cdots, -1, -1, \cdots, -1}^{N})~,\notag \\
& \tilde{I}_2 \equiv{\mbox{diag}}(+1, \cdots,+1, -1, \cdots, -1, +1, \cdots, +1,-1, \cdots, -1)~, \notag \\
& \tilde{I}_3 \equiv
{\mbox{diag}}(\underbrace{+1, \cdots,+1}_{p}, \underbrace{-1, \cdots, -1}_{q}, \underbrace{-1, \cdots, -1}_{r},
\underbrace{+1, \cdots, +1}_{s(=N-p-q-r)})~,
\label{tildeI}
\end{align}
where $p,q,r,s\geq 0$ and $N=p+q+r+s$.
We denote the above BC (\ref{typeI}) as $[p,q;r,s]^{\mathrm{I}}$.
The symmetry of $[p,q;r,s]^{\mathrm{I}}$ becomes
\begin{align}
SO(2N) \rightarrow SO(2p) \times SO(2q) \times SO(2r) \times SO(2s)~,
\label{I-break}
\end{align}
where $SO(0)$ means nothing.

~~\\
(Type-II)~~Two of them belong to $P^{(2)}$ type and a remaining one is $P^{(0)}$ type, 
and they are classified into the three subtypes:
\begin{align}
(P_0 ,P_1 ,U)
&=(\sigma_0 \otimes \tilde{I}_1, \sigma_2 \otimes \tilde{I}_2, \sigma_2 \otimes \tilde{I}_3)~~~(\mbox{Type-IIa})~,
\label{IIa}\\
&=(\sigma_2 \otimes \tilde{I}_1, \sigma_0 \otimes \tilde{I}_2, \sigma_2 \otimes \tilde{I}_3)~~~(\mbox{Type-IIb})~,
\label{IIb}\\
&=(\sigma_2 \otimes \tilde{I}_1, \sigma_2 \otimes \tilde{I}_2, \sigma_0 \otimes \tilde{I}_3)~~~(\mbox{Type-IIc})~,
\label{IIc}
\end{align}
where $\tilde{I}_1$, $\tilde{I}_2$ and $\tilde{I}_3$ are defined by (\ref{tildeI}).
We denote the above BCs (\ref{IIa}), (\ref{IIb}) and (\ref{IIc}) as $[p,q;r,s]^{\mathrm{IIa}}$, $[p,q;r,s]^{\mathrm{IIb}}$
and $[p,q;r,s]^{\mathrm{IIc}}$, respectively.
The symmetries of $[p,q;r,s]^{\mathrm{IIa}}$, $[p,q;r,s]^{\mathrm{IIb}}$ and $[p,q;r,s]^{\mathrm{IIc}}$ become
\begin{align}
& SO(2N) \rightarrow SU(p+q) \times SU(r+s) \times U(1)^{2-k}~~~(\mbox{Type-IIa})~,
\label{IIa-break}\\
& SO(2N) \rightarrow SU(p+r) \times SU(q+s) \times U(1)^{2-k}~~~(\mbox{Type-IIb})~,
\label{IIb-break}\\
& SO(2N) \rightarrow SU(p+s) \times SU(q+r) \times U(1)^{2-k}~~~(\mbox{Type-IIc})~,
\label{IIc-break}
\end{align}
where $k$ is a sum of the number of $SU(0)$ and $SU(1)$,  $SU(0)$ means nothing and $SU(1)$ unconventionally stands for $U(1)$.
Because Type-IIa, Type-IIb and Type-IIc are interchanged among them by the interchange of $P_0$, $P_1$ and $U$
and the same results for numbers of each species are obtained,
we use type-IIa as the representative of type-II.
If two BCs are transformed into each other by a global $SO(2N)$ transformation and/or a gauge transformation,
they are equivalent.
The $[p,q;r,s]^{\mathrm{IIa}}$ is transformed into $[p + \ell_1 , q - \ell_1, r + \ell_2, s - \ell_2]^{\mathrm{IIa}}$
using the global $SO(2N)$ symmetry which changes $\sigma_2$ into $-\sigma_2$ partially.
Here, $\ell_1$ and $\ell_2$ are arbitrary integers which satisfy $p + \ell_1 , q - \ell_1, r + \ell_2, s - \ell_2 \ge 0$.
Hence we use $[m, 0, n, 0]^{\mathrm{IIa}}$ $(N=m+n)$ in place of $[p,q;r,s]^{\mathrm{IIa}}$ with $m = p+q$ and $n=r+s$.
In the appendix A, we discuss the gauge invariance of BCs and the equivalence relations for the sake of completeness.

Strictly speaking, we must find the minimum of the effective potential for the Wilson line phases
in order to know physical gauge symmetry~\cite{H}.
It requires a model-dependent analysis because the effective potential depends on the particle contents and their BCs.
In the following discussion, we suppose that  
the BC belongs to the same equivalence class of $(P_0^{\rm sym}, P_1^{\rm sym}, U^{\rm sym})$ defined by (\ref{equiv3}).

\subsection{$Z_2$ parity assignment}

We study the $Z_2$ parity assignment for gauge fields and matter fermions for two types.

~~\\
(Type-I)~~The BCs of gauge fields, $A_M(x,y) = A^{\alpha}_M(x,y)T^{\alpha}$, are given by
\begin{align}
& s_0:~A_{\mu}(x,-y)=P_0A_{\mu}(x,y)P_0^{-1}~,~~ A_{y}(x,-y)=-P_0A_{y}(x,y)P_0^{-1}~,  \notag \\
& s_1:~A_{\mu}(x,2\pi R-y)=P_1A_{\mu}(x,y)P_1^{-1}~,~~ A_{y}(x,2\pi R-y)=-P_1A_{y}(x,y)P_1^{-1}~, \notag \\ 
& t:~A_{M}(x,y+2\pi R)=UA_{M}(x,y)U^{-1}~,
\label{gauge-BCs}
\end{align}
where $M=0, \cdots, 3, 5$.
Using the relation ${\rm tr}(T^{\alpha} T^{\beta}) = \delta^{\alpha\beta}/2$,
the BCs for four-dimensional components of gauge bosons, $A_{\mu}(x,y) = A^{\alpha}_{\mu}(x,y)T^{\alpha}$, are rewritten as
\begin{align}
&A_{\mu}^{\alpha}(x,-y)=2{\rm tr}(T^{\alpha}P_0T^{\beta}P_0^{-1})A_{\mu}^{\beta}(x,y)~,~~
A_{\mu}^{\alpha}(x,2\pi R-y)=2{\rm tr}(T^{\alpha}P_1T^{\beta}P_1^{-1})A_{\mu}^{\beta}(x,y)~,~~
\notag\\
&A_{\mu}^{\alpha}(x,y+2\pi R)=2{\rm tr}(T^{\alpha}UT^{\beta}U^{-1})A_{\mu}^{\beta}(x,y)~.
\label{gauge-BCs2}
\end{align}
Under the BC $[p, q, r, s]^{\mathrm{I}}$, $A_{\mu}^{\alpha}$ is decomposed into a sum of multiplets of the subgroup
$SO(2N) \rightarrow SO(2p) \times SO(2q) \times SO(2r) \times SO(2s)$ $(N=p+q+r+s)$ as
\begin{align}
& \mathbf{N(2N-1)} = (\mathbf{p(2p-1)},\mathbf{1},\mathbf{1},\mathbf{1})^{++;+}
+(\mathbf{1},\mathbf{q(2q-1)},\mathbf{1},\mathbf{1})^{++;+} \notag\\
&~~ +(\mathbf{1},\mathbf{1},\mathbf{r(2r-1)},\mathbf{1})^{++;+} 
+(\mathbf{1},\mathbf{1},\mathbf{1},\mathbf{s(2s-1)})^{++;+} \notag\\
&~~ +(\mathbf{2p},\mathbf{2q},\mathbf{1},\mathbf{1})^{+-;-}
+(\mathbf{2p},\mathbf{1},\mathbf{2r},\mathbf{1})^{-+;-}
+(\mathbf{2p},\mathbf{1},\mathbf{1},\mathbf{2s})^{--;+} \notag\\
&~~ +(\mathbf{1},\mathbf{2q},\mathbf{2r},\mathbf{1})^{--;+}
+(\mathbf{1},\mathbf{2q},\mathbf{1},\mathbf{2s})^{-+;-}
+(\mathbf{1},\mathbf{1},\mathbf{2r},\mathbf{2s})^{+-;-}~,
\label{SOGparity}
\end{align}
where $Z_2$ parities are obtained using the formulea (\ref{gauge-BCs2}),
and $\mathbf{p(2p-1)}$ and $\mathbf{2p}$ represent the components of $A_{\mu}^{\alpha}$ 
with adjoint and vector representation of $SO(2p)$, respectively.
The index $+$ or $-$ stands for $Z_2$ parity $+1$ or $-1$.
The $A_{y}^{\alpha}$ have the opposite $Z_2$ parities $\mathcal{P}_0$ and $\mathcal{P}_1$ to those of $A_{\mu}^{\alpha}$.

We require the $Z_2$ parity invariance for the interaction between the gauge fields and a matter fermion $\psi$:
\begin{align}
\mathcal{P}_0(\bar{\psi}\gamma^M A_M^{\alpha} T^{\alpha} \psi) 
= \mathcal{P}_1(\bar{\psi}\gamma^M A_M^{\alpha} T^{\alpha} \psi) =+1~. 
\label{bp-gA-p}
\end{align}
The invariance under the shift $y \to y + 2\pi R$, i.e., 
$\mathcal{U}(\bar{\psi}\gamma^M A_M^{\alpha} T^{\alpha} \psi) =+1$, is automatically satisfied 
from $\mathcal{P}_0 \mathcal{P}_1 \mathcal{U} = +1$.

There are two inequivalent spinor representations ${\mathbf{2}}^{N-1}_1$ and ${\mathbf{2}}^{N-1}_2$ in $SO(2N)$.
For $N=4\ell+1$ and $4\ell+3$ $(\ell \in \{\mathbb{N},0\})$,
$\mathbf{2}^{N-1}_a$ are complex representations and they are conjugate to each other, i.e.,
$\overline{{\mathbf{2}}}^{N-1}_1 = {\mathbf{2}}^{N-1}_2$ and $\overline{{\mathbf{2}}}^{N-1}_2 = {\mathbf{2}}^{N-1}_1$.
For $N=4\ell$ $(\ell \in \mathbb{N})$, 
$\mathbf{2}^{N-1}_a$ $(a=1,2)$ are real representations and self-conjugate, i.e., $\overline{\mathbf{2}}_{a}^{N-1}=\mathbf{2}^{N-1}_a$.
For $4\ell+2$ $(\ell \in \{\mathbb{N},0\})$, 
$\mathbf{2}^{N-1}_a$ are pseudo real representations and self-conjugate.
If the matter fermion forms the spinor representations ${\mathbf{2}}^{N-1}_a$ or
the vector representation $\mathbf{2N}$, the following relations hold:
\begin{align}
&\mathcal{P}_0(\overline{\mathbf{2}}_{a}^{N-1} \times \mathbf{N(2N-1)} \times \mathbf{2}^{N-1}_a )=
\mathcal{P}_1(\overline{\mathbf{2}}_{a}^{N-1} \times \mathbf{N(2N-1)} \times \mathbf{2}^{N-1}_a )=+1~,
\label{2^N-1condition}\\
&\mathcal{P}_0({\mathbf{2N}} \times \mathbf{N(2N-1)} \times \mathbf{2N} )=
\mathcal{P}_1({\mathbf{2N}} \times \mathbf{N(2N-1)} \times \mathbf{2N} )=+1~.
\label{Ncondition}
\end{align}

By the $Z_2$ projection with $P_0$, $SO(2N)$ is broken down to $SO(2(p+q)) \times SO(2(r+s))$ 
and $\mathbf{2}^{N-1}_a$ and $\mathbf{2N}$ are decomposed into
\begin{align}
&\mathbf{2}_1^{N-1}=(\mathbf{2}_1^{p+q-1},\mathbf{2}_1^{r+s-1})+(\mathbf{2}_2^{p+q-1},\mathbf{2}_2^{r+s-1})~,~~ 
\mathbf{2}_2^{N-1}=(\mathbf{2}_1^{p+q-1},\mathbf{2}_2^{r+s-1})+(\mathbf{2}_2^{p+q-1},\mathbf{2}_1^{r+s-1})~, \notag \\
&\mathbf{2N}=(\mathbf{2(p+q)},\mathbf{1})+(\mathbf{1},\mathbf{2(r+s)})~.
\end{align}
Using  (\ref{SOGparity}), (\ref{2^N-1condition}) and  (\ref{Ncondition}), 
we find that each multiplet has a definite $\mathcal{P}_0$ as
\begin{align}
&\mathcal{P}_0 ((\mathbf{2}_1^{p+q-1},\mathbf{2}_1^{r+s-1}))=+\eta_1^0~,~~
\mathcal{P}_0 ((\mathbf{2}_2^{p+q-1},\mathbf{2}_2^{r+s-1}))=-\eta_1^0~, \notag \\
&\mathcal{P}_0 ((\mathbf{2}_1^{p+q-1},\mathbf{2}_2^{r+s-1}))=+\eta_2^0~,~~
\mathcal{P}_0 ((\mathbf{2}_2^{p+q-1},\mathbf{2}_1^{r+s-1}))=-\eta_2^0~, \notag \\
&\mathcal{P}_0 ((\mathbf{2(p+q)},\mathbf{1}))=+\eta_{\rm{v}}^0~,~~
\mathcal{P}_0 ((\mathbf{1},\mathbf{2(r+s)}))=-\eta_{\rm{v}}^0~,
\label{P0Z2type1}
\end{align}
where $\eta_1^0$, $\eta_2^0$ and $\eta_{\rm{v}}^0$ are intrinsic $Z_2$ parities.
In the same way, $SO(2N)$ is broken down to $SO(2(p+r)) \times SO(2(q+s))$ by $P_1$,
and $\mathbf{2}^{N-1}_a$ and $\mathbf{2N}$ are decomposed into
\begin{align}
&\mathbf{2}_1^{N-1}=(\mathbf{2}_1^{p+r-1},\mathbf{2}_1^{q+s-1})+(\mathbf{2}_2^{p+r-1},\mathbf{2}_2^{q+s-1})~,~~
\mathbf{2}_2^{N-1}=(\mathbf{2}_1^{p+r-1},\mathbf{2}_2^{q+s-1})+(\mathbf{2}_2^{p+r-1},\mathbf{2}_1^{q+s-1})~, \notag \\
&\mathbf{2N}=(\mathbf{2(p+r)},\mathbf{1})+(\mathbf{1},\mathbf{2(q+s)})~.
\end{align}
Each multiplet has a definite $\mathcal{P}_1$ as
\begin{align}
&\mathcal{P}_1 ((\mathbf{2}_1^{p+r-1},\mathbf{2}_1^{q+s-1}))=+\eta_1^1~,~~
\mathcal{P}_1 ((\mathbf{2}_2^{p+r-1},\mathbf{2}_2^{q+s-1}))=-\eta_1^1~, \notag \\
&\mathcal{P}_1 ((\mathbf{2}_1^{p+r-1},\mathbf{2}_2^{q+s-1}))=+\eta_2^1~,~~
\mathcal{P}_1 ((\mathbf{2}_2^{p+r-1},\mathbf{2}_1^{q+s-1}))=-\eta_2^1~, \notag \\
&\mathcal{P}_1 ((\mathbf{2(p+r)},\mathbf{1}))=+\eta_{\rm{v}}^1~,~~
\mathcal{P}_1 ((\mathbf{1},\mathbf{2(q+s)}))=-\eta_{\rm{v}}^1~,
\label{P1Z2type1}
\end{align}
where $\eta_1^1$, $\eta_2^1$ and $\eta_{\rm{v}}^1$ are intrinsic $Z_2$ parities.
The same argument holds for $U$.

Combining the $Z_2$ projections with $P_0$ and $P_1$,
$SO(2N)$ is broken down to $SO(2p) \times SO(2q) \times SO(2r) \times SO(2s)$,
and $\mathbf{2}^{N-1}_a$ and $\mathbf{2N}$ are decomposed into
\begin{eqnarray}
&~& \mathbf{2}_1^{N-1}
=(\mathbf{2}_1^{p-1},\mathbf{2}_1^{q-1},\mathbf{2}_1^{r-1},\mathbf{2}_1^{s-1})
+(\mathbf{2}_1^{p-1},\mathbf{2}_1^{q-1},\mathbf{2}_2^{r-1},\mathbf{2}_2^{s-1}) \notag \\
&~& \hspace{20mm}+(\mathbf{2}_2^{p-1},\mathbf{2}_2^{q-1},\mathbf{2}_1^{r-1},\mathbf{2}_1^{s-1})
+(\mathbf{2}_2^{p-1},\mathbf{2}_2^{q-1},\mathbf{2}_2^{r-1},\mathbf{2}_2^{s-1}) \notag \\
&~& \hspace{20mm}+(\mathbf{2}_1^{p-1},\mathbf{2}_2^{q-1},\mathbf{2}_1^{r-1},\mathbf{2}_2^{s-1})
+(\mathbf{2}_1^{p-1},\mathbf{2}_2^{q-1},\mathbf{2}_2^{r-1},\mathbf{2}_1^{s-1}) \notag \\
&~& \hspace{20mm}+(\mathbf{2}_2^{p-1},\mathbf{2}_1^{q-1},\mathbf{2}_1^{r-1},\mathbf{2}_2^{s-1})
+(\mathbf{2}_2^{p-1},\mathbf{2}_1^{q-1},\mathbf{2}_2^{r-1},\mathbf{2}_1^{s-1})~, \notag \\ 
&~& \mathbf{2}_2^{N-1}
=(\mathbf{2}_1^{p-1},\mathbf{2}_1^{q-1},\mathbf{2}_1^{r-1},\mathbf{2}_2^{s-1})
+(\mathbf{2}_1^{p-1},\mathbf{2}_1^{q-1},\mathbf{2}_2^{r-1},\mathbf{2}_1^{s-1}) \notag \\
&~& \hspace{20mm}+(\mathbf{2}_2^{p-1},\mathbf{2}_2^{q-1},\mathbf{2}_1^{r-1},\mathbf{2}_2^{s-1})
+(\mathbf{2}_2^{p-1},\mathbf{2}_2^{q-1},\mathbf{2}_2^{r-1},\mathbf{2}_1^{s-1}) \notag \\
&~& \hspace{20mm}+(\mathbf{2}_1^{p-1},\mathbf{2}_2^{q-1},\mathbf{2}_1^{r-1},\mathbf{2}_1^{s-1})
+(\mathbf{2}_1^{p-1},\mathbf{2}_2^{q-1},\mathbf{2}_2^{r-1},\mathbf{2}_2^{s-1}) \notag \\
&~& \hspace{20mm}+(\mathbf{2}_2^{p-1},\mathbf{2}_1^{q-1},\mathbf{2}_1^{r-1},\mathbf{2}_1^{s-1})
+(\mathbf{2}_2^{p-1},\mathbf{2}_1^{q-1},\mathbf{2}_2^{r-1},\mathbf{2}_2^{s-1})~,\notag \\
&~& \mathbf{2N}
=(\mathbf{2p},\mathbf{1},\mathbf{1},\mathbf{1})
+(\mathbf{1},\mathbf{2q},\mathbf{1},\mathbf{1})
+(\mathbf{1},\mathbf{1},\mathbf{2r},\mathbf{1})
+(\mathbf{1},\mathbf{1},\mathbf{1},\mathbf{2s})~.
\end{eqnarray}
The $Z_2$ parities of each multiplet are lised in Table \ref{pqrs}.
\begin{table}[htbp]
\begin{center}
\begin{tabular}{c|c|c|c|c}
\hline
\multicolumn{2}{c|}{Representation}&$\mathcal{P}_0$&$\mathcal{P}_1$&$\mathcal{U}$\\ \hline \hline
$\mathbf{2}_1^{N-1}$&$(\mathbf{2}_1^{p-1},\mathbf{2}_1^{q-1},\mathbf{2}_1^{r-1},\mathbf{2}_1^{s-1})$&$+\eta_1^0$ &$+\eta_1^1$ &$+\eta_1^0\eta_1^1$ \\
&$(\mathbf{2}_1^{p-1},\mathbf{2}_1^{q-1},\mathbf{2}_2^{r-1},\mathbf{2}_2^{s-1})$&$+\eta_1^0$ &$-\eta_1^1$ &$-\eta_1^0\eta_1^1$ \\
&$(\mathbf{2}_2^{p-1},\mathbf{2}_2^{q-1},\mathbf{2}_1^{r-1},\mathbf{2}_1^{s-1})$&$+\eta_1^0$ &$-\eta_1^1$ &$-\eta_1^0\eta_1^1$ \\
&$(\mathbf{2}_2^{p-1},\mathbf{2}_2^{q-1},\mathbf{2}_2^{r-1},\mathbf{2}_2^{s-1})$&$+\eta_1^0$ &$+\eta_1^1$ &$+\eta_1^0\eta_1^1$ \\
&$(\mathbf{2}_1^{p-1},\mathbf{2}_2^{q-1},\mathbf{2}_1^{r-1},\mathbf{2}_2^{s-1})$&$-\eta_1^0$ &$+\eta_1^1$ &$-\eta_1^0\eta_1^1$ \\
&$(\mathbf{2}_1^{p-1},\mathbf{2}_2^{q-1},\mathbf{2}_2^{r-1},\mathbf{2}_1^{s-1})$&$-\eta_1^0$ &$-\eta_1^1$ &$+\eta_1^0\eta_1^1$ \\
&$(\mathbf{2}_2^{p-1},\mathbf{2}_1^{q-1},\mathbf{2}_1^{r-1},\mathbf{2}_2^{s-1})$&$-\eta_1^0$ &$-\eta_1^1$ &$+\eta_1^0\eta_1^1$ \\
&$(\mathbf{2}_2^{p-1},\mathbf{2}_1^{q-1},\mathbf{2}_2^{r-1},\mathbf{2}_1^{s-1})$&$-\eta_1^0$ &$+\eta_1^1$ &$-\eta_1^0\eta_1^1$ \\ 
\hline 
$\mathbf{2}_2^{N-1}$&
$(\mathbf{2}_1^{p-1},\mathbf{2}_1^{q-1},\mathbf{2}_1^{r-1},\mathbf{2}_2^{s-1})$&$+\eta_2^0$ &$+\eta_2^1$ &$+\eta_2^0\eta_2^1$ \\
&$(\mathbf{2}_1^{p-1},\mathbf{2}_1^{q-1},\mathbf{2}_2^{r-1},\mathbf{2}_1^{s-1})$&$+\eta_2^0$ &$-\eta_2^1$ &$-\eta_2^0\eta_2^1$ \\
&$(\mathbf{2}_2^{p-1},\mathbf{2}_2^{q-1},\mathbf{2}_1^{r-1},\mathbf{2}_2^{s-1})$&$+\eta_2^0$ &$-\eta_2^1$ &$-\eta_2^0\eta_2^1$ \\
&$(\mathbf{2}_2^{p-1},\mathbf{2}_2^{q-1},\mathbf{2}_2^{r-1},\mathbf{2}_1^{s-1})$&$+\eta_2^0$ &$+\eta_2^1$ &$+\eta_2^0\eta_2^1$ \\
&$(\mathbf{2}_1^{p-1},\mathbf{2}_2^{q-1},\mathbf{2}_1^{r-1},\mathbf{2}_1^{s-1})$&$-\eta_2^0$ &$+\eta_2^1$ &$-\eta_2^0\eta_2^1$ \\
&$(\mathbf{2}_1^{p-1},\mathbf{2}_2^{q-1},\mathbf{2}_2^{r-1},\mathbf{2}_2^{s-1})$&$-\eta_2^0$ &$-\eta_2^1$ &$+\eta_2^0\eta_2^1$ \\
&$(\mathbf{2}_2^{p-1},\mathbf{2}_1^{q-1},\mathbf{2}_1^{r-1},\mathbf{2}_1^{s-1})$&$-\eta_2^0$ &$-\eta_2^1$ &$+\eta_2^0\eta_2^1$ \\
&$(\mathbf{2}_2^{p-1},\mathbf{2}_1^{q-1},\mathbf{2}_2^{r-1},\mathbf{2}_2^{s-1})$&$-\eta_2^0$ &$+\eta_2^1$ &$-\eta_2^0\eta_2^1$ \\
\hline
$\mathbf{2N}$&
$(\mathbf{2p},\mathbf{1},\mathbf{1},\mathbf{1})$&
$+\eta_{\rm{v}}^0$ &$+\eta_{\rm{v}}^1$ &$+\eta_{\rm{v}}^0\eta_{\rm{v}}^1$ \\
&$(\mathbf{1},\mathbf{2q},\mathbf{1},\mathbf{1})$&
$+\eta_{\rm{v}}^0$ &$-\eta_{\rm{v}}^1$ &$-\eta_{\rm{v}}^0\eta_{\rm{v}}^1$ \\
&$(\mathbf{1},\mathbf{1},\mathbf{2r},\mathbf{1})$&
$-\eta_{\rm{v}}^0$ &$+\eta_{\rm{v}}^1$ &$-\eta_{\rm{v}}^0\eta_{\rm{v}}^1$ \\
&$(\mathbf{1},\mathbf{1},\mathbf{1},\mathbf{2s})$&
$-\eta_{\rm{v}}^0$ &$-\eta_{\rm{v}}^1$ &$+\eta_{\rm{v}}^0\eta_{\rm{v}}^1$ \\
\hline
\end{tabular}
\end{center}
\caption{$Z_2$ parities of matter fermions in Type-I.}
\label{pqrs}
\end{table}
The eigenvalues of $\mathcal{U}$ are determined from $\mathcal{P}_0 \mathcal{P}_1 \mathcal{U} = +1$.

In the case with $s=0$, $\mathbf{2}^{N-1}_a$ and $\mathbf{2N}$ are decomposed into
\begin{align}
&\mathbf{2}_1^{N-1}
=(\mathbf{2}_1^{p-1},\mathbf{2}_1^{q-1},\mathbf{2}_1^{r-1})
+(\mathbf{2}_2^{p-1},\mathbf{2}_2^{q-1},\mathbf{2}_1^{r-1}) 
+(\mathbf{2}_1^{p-1},\mathbf{2}_2^{q-1},\mathbf{2}_2^{r-1})
+(\mathbf{2}_2^{p-1},\mathbf{2}_1^{q-1},\mathbf{2}_2^{r-1})~, \notag \\ 
&\mathbf{2}_2^{N-1}
=(\mathbf{2}_1^{p-1},\mathbf{2}_1^{q-1},\mathbf{2}_2^{r-1})
+(\mathbf{2}_1^{p-1},\mathbf{2}_2^{q-1},\mathbf{2}_1^{r-1}) 
+(\mathbf{2}_2^{p-1},\mathbf{2}_1^{q-1},\mathbf{2}_1^{r-1})
+(\mathbf{2}_2^{p-1},\mathbf{2}_2^{q-1},\mathbf{2}_2^{r-1})~,\notag \\
&\mathbf{2N}
=(\mathbf{2p},\mathbf{1},\mathbf{1})
+(\mathbf{1},\mathbf{2q},\mathbf{1})
+(\mathbf{1},\mathbf{1},\mathbf{2r})~,
\end{align}
under $SO(2p) \times SO(2q) \times SO(2r)$.
In the case with $r=s=0$, $\mathbf{2}^{N-1}_a$ and $\mathbf{2N}$ are decomposed into
\begin{align}
&\mathbf{2}_1^{N-1}
=(\mathbf{2}_1^{p-1},\mathbf{2}_1^{q-1})+(\mathbf{2}_2^{p-1},\mathbf{2}_2^{q-1})~,~~ 
\mathbf{2}_2^{N-1}
=(\mathbf{2}_1^{p-1},\mathbf{2}_2^{q-1})+(\mathbf{2}_2^{p-1},\mathbf{2}_1^{q-1})~,\notag \\
&\mathbf{2N}
=(\mathbf{2p},\mathbf{1})
+(\mathbf{1},\mathbf{2q})~,
\end{align}
under $SO(2p) \times SO(2q)$.
The $Z_2$ parities of each multiplet are understood from those for the corresponding representations in Table \ref{pqrs}.

~~\\
(Type-II)~~Under the BC $[m, 0, n, 0]^{\mathrm{IIa}}$, $A_{\mu}^{\alpha}$ is decomposed into a sum of multiplets of the subgroup
$SO(2N) \rightarrow SU(m) \times SU(n) \times U(1)^{2-k}$ $(N=m+n)$ as
\begin{align}
& \mathbf{N(2N-1)} = (\mathbf{m^2-1},\mathbf{1})^{++;+}+(\mathbf{1},\mathbf{n^2-1})^{++;+}
+(\mathbf{1},\mathbf{1})^{++;+}+(\mathbf{1},\mathbf{1})^{++;+}
\notag\\
&~~ + (\mathbf{m},\overline{\mathbf{n}})^{-+;-} + (\overline{\mathbf{m}},\mathbf{n})^{-+;-}
+ \left(\mathbf{m(m-1)/2},\mathbf{1}\right)^{+-;-} 
+ \left(\mathbf{1},\mathbf{n(n-1)/2}\right)^{+-;-}
\notag\\
&~~ + \left(\overline{\mathbf{m(m-1)/2}},\mathbf{1}\right)^{+-;-} 
+ \left(\mathbf{1},\overline{\mathbf{n(n-1)/2}}\right)^{+-;-}
+ (\mathbf{m},\mathbf{n})^{--;+}
+ (\overline{\mathbf{m}},\overline{\mathbf{n}})^{--;+}~,
\label{SOGparity2}
\end{align}
where $Z_2$ parities are obtained using the formulea (\ref{gauge-BCs2}), $U(1)$ charges are omitted, and
$\mathbf{m^2-1}$ $(\mathbf{n^2-1})$, $\mathbf{m}$ $(\mathbf{n})$ 
and $\mathbf{m(m-1)/2}$ $(\mathbf{n(n-1)/2})$ represent the components of $A_{\mu}^{\alpha}$ 
with  adjoint, vector and rank 2 antisymmetric representation of $SU(m)$ $(SU(n))$, respectively.
The representations with overline stand for the complex conjugate ones.

By the $Z_2$ projection with $P_1$, $SO(2N)$ is broken down to its subgroup including $SU(N)$ 
whose adjoint representation $\mathbf{N^2-1}$ is given by
\begin{align}
\mathbf{N^2-1} =&(\mathbf{m^2-1},\mathbf{1})^{++;+}+(\mathbf{1},\mathbf{n^2-1})^{++;+}
+(\mathbf{1},\mathbf{1})^{++;+}
\notag\\
&+(\mathbf{m},\overline{\mathbf{n}})^{-+;-} + (\overline{\mathbf{m}},\mathbf{n})^{-+;-}~.
\label{AdNGG}
\end{align}
In the same way, by $U$, $SO(2N)$ is broken down to its subgroup including $SU(N)$ 
whose adjoint representation $\mathbf{N^2-1}$ is given by
\begin{align}
\mathbf{N^2-1}=&(\mathbf{m^2-1},\mathbf{1})^{++;+}+(\mathbf{1},\mathbf{n^2-1})^{++;+}
+(\mathbf{1},\mathbf{1})^{++;+}
\notag\\
& + (\mathbf{m},\mathbf{n})^{--;+}
+ (\overline{\mathbf{m}},\overline{\mathbf{n}})^{--;+}~.
\label{AdNFl}
\end{align}
Under the exchange of $P_1$ and $U$, the adjoint representations (\ref{AdNGG}) and (\ref{AdNFl}) are exchanged.
It corresponds to the relation between Georgi-Glashow type $SU(5)$~\cite{GG} and the Flipped type $SU(5)$~\cite{Flipped}
in $SO(10)$ GUTs~\cite{SO10,LR}.

We study the $Z_2$ parity assignment for matter fermions.
By the $Z_2$ projection with $P_0$, $SO(2N)$ is broken down to $SO(2m) \times SO(2n)$,
and $\mathbf{2}^{N-1}_a$ and $\mathbf{2N}$ are decomposed into
\begin{align}
&\mathbf{2}_1^{N-1}=(\mathbf{2}_1^{m-1},\mathbf{2}_1^{n-1})+(\mathbf{2}_2^{m-1},\mathbf{2}_2^{n-1})~,~~ 
\mathbf{2}_2^{N-1}=(\mathbf{2}_1^{m-1},\mathbf{2}_2^{n-1})+(\mathbf{2}_2^{m-1},\mathbf{2}_1^{n-1})~, \notag \\
&\mathbf{2N}=(\mathbf{2m},\mathbf{1})+(\mathbf{1},\mathbf{2n})~.
\end{align}
Using  (\ref{SOGparity}), (\ref{2^N-1condition}) and  (\ref{Ncondition}), 
we find that each multiplet has a definite $\mathcal{P}_0$ as
\begin{align}
&\mathcal{P}_0((\mathbf{2}_1^{m-1},\mathbf{2}_1^{n-1}))=+\eta_1^0~,~~\mathcal{P}_0((\mathbf{2}_2^{m-1},\mathbf{2}_2^{n-1}))=-\eta_1^0~,\notag \\
&\mathcal{P}_0((\mathbf{2}_1^{m-1},\mathbf{2}_2^{n-1}))=+\eta_2^0~,~~\mathcal{P}_0((\mathbf{2}_2^{m-1},\mathbf{2}_1^{n-1}))=-\eta_2^0~,\notag \\
&\mathcal{P}_0((\mathbf{2m},\mathbf{1}))=+\eta_{\rm{v}}^0~,~~\mathcal{P}_0((\mathbf{1},\mathbf{2n}))=-\eta_{\rm{v}}^0~.
\end{align}
In the same way, $SO(2N)$ is broken down to $SU(N) \times U(1)$ by $P_1$,
and $\mathbf{2}^{N-1}_a$ and $\mathbf{2N}$ are decomposed into
\begin{align}
\mathbf{2}_1^{N-1}=\sum_{k=0}^{[N/2]}[N,2k]~,~~ \mathbf{2}_2^{N-1}=\sum_{k=0}^{[(N-1)/2]}[N,2k+1]~,~~ 
\mathbf{2N}=\mathbf{N}+\overline{\mathbf{N}}~,
\end{align}
where $[N/2]$ and $[(N-1)/2]$ represent Gauss's symbol, and $[N,2k](={}_NC_{2k})$ and $[N,2k+1](={}_NC_{2k+1})$ 
are the rank $2k + 1$ totally antisymmetric representations of $SU(N)$ gauge group and the $U(1)$ charge is omitted.  
Each multiplet has a definite $\mathcal{P}_1$ as
\begin{align}
& \mathcal{P}_1([N,2k])=(-1)^{k}\eta_1^1~,~~\mathcal{P}_1([N,2k+1])=(-1)^{k}\eta_2^1~,~~ \notag \\
& \mathcal{P}_1(\mathbf{N})=+\eta_{\rm{v}}^1~,~~\mathcal{P}_1(\overline{\mathbf{N}})=-\eta_{\rm{v}}^1~.
\end{align}
The same argument holds for $U$.

Combining the $Z_2$ projections with $P_0$ and $P_1$,
$SO(2N)$ is broken down to $SU(m) \times SU(n) \times U(1)^{2-k}$,
and $\mathbf{2}^{N-1}_a$ and $\mathbf{2N}$ are decomposed into
\begin{eqnarray}
&~& \mathbf{2}_1^{N-1}
=\sum_{k=0}^{[N/2]}\sum_{\ell=0}^{2k}({}_mC_{\ell},{}_nC_{2k-\ell}) \notag \\
&~& \hspace{8.5mm}
=\sum_{k=0}^{[N/2]}\sum_{\ell=\mathrm{even}}({}_mC_{\ell},{}_nC_{2k-\ell})
+\sum_{k=0}^{[N/2]}\sum_{\ell=\mathrm{odd}}({}_mC_{\ell},{}_nC_{2k-\ell})~,\notag \\
&~& \mathbf{2}_2^{N-1}
=\sum_{k=0}^{[(N-1)/2]}\sum_{\ell=0}^{2k+1}({}_mC_{\ell},{}_nC_{2k-\ell+1}) \notag \\
&~& \hspace{8.5mm}=\sum_{k=0}^{[(N-1)/2]}\sum_{\ell=\mathrm{even}}({}_mC_{\ell},{}_nC_{2k-\ell+1})
+\sum_{k=0}^{[(N-1)/2]}\sum_{\ell=\mathrm{odd}}({}_mC_{\ell},{}_nC_{2k-\ell+1})~, \notag \\
&~& \mathbf{2N}=(\mathbf{m},\mathbf{1})+(\mathbf{1},\mathbf{n})+(\overline{\mathbf{m}},\mathbf{1})+(\mathbf{1},\overline{\mathbf{n}})~.
\end{eqnarray}
The $Z_2$ parities of each multiplet are listed in Table \ref{IIaZ2}.
\begin{table}[htbp]
\begin{center}
\begin{tabular}{c|c|c|c|c}
\hline
\multicolumn{2}{c|}{Representation}&$\mathcal{P}_0$&$\mathcal{P}_1$&$\mathcal{U}$ \\ \hline \hline
$\mathbf{2}_1^{N-1}$&$\textstyle({}_mC_{\ell},{}_nC_{2k-\ell})_{\ell=\mathrm{even}}$&$+\eta_1^0$&$(-1)^k\eta_1^1$&$(-1)^{k}\eta_1^0\eta_1^1$ \\
 &$\textstyle({}_mC_{\ell},{}_nC_{2k-\ell})_{\ell=\mathrm{odd}}$&$-\eta_1^0$&$(-1)^k\eta_1^1$&$-(-1)^{k}\eta_1^0\eta_1^1$ \\ \hline
$\mathbf{2}_2^{N-1}$&$\textstyle({}_mC_{\ell},{}_nC_{2k-\ell+1})_{\ell=\mathrm{even}}$&$+\eta_2^0$&$(-1)^k\eta_2^1$&$(-1)^{k}\eta_2^0\eta_2^1$ \\
 &$\textstyle({}_mC_{\ell},{}_nC_{2k-\ell+1})_{\ell=\mathrm{odd}}$&$-\eta_2^0$&$(-1)^k\eta_2^1$&$-(-1)^{k}\eta_2^0\eta_2^1$ \\ \hline
$\mathbf{2N}$&$(\mathbf{m},\mathbf{1})$&$+\eta_{\rm{v}}^0$&$+\eta_{\rm{v}}^1$&$+\eta_{\rm{v}}^0\eta_{\rm{v}}^1$ \\
&$(\mathbf{1},\mathbf{n})$&$-\eta_{\rm{v}}^0$&$+\eta_{\rm{v}}^1$&$-\eta_{\rm{v}}^0\eta_{\rm{v}}^1$ \\
&$(\overline{\mathbf{m}},\mathbf{1})$&$+\eta_{\rm{v}}^0$&$-\eta_{\rm{v}}^1$&$-\eta_{\rm{v}}^0\eta_{\rm{v}}^1$ \\
&$(\mathbf{1},\overline{\mathbf{n}})$&$-\eta_{\rm{v}}^0$&$-\eta_{\rm{v}}^1$&$+\eta_{\rm{v}}^0\eta_{\rm{v}}^1$ \\ \hline
\end{tabular}
\end{center}
\caption{$Z_2$ parities of matter fermions in Type-IIa.}
\label{IIaZ2}
\end{table}
The $Z_2$ parity assignments for Type-IIb and Type-IIc are obtained 
by the exchange of $\mathcal{P}_0$, $\mathcal{P}_1$ and $\mathcal{U}$, i.e.,
$(\mathcal{P}_0, \mathcal{P}_1, \mathcal{U})^{\mathrm{IIa}} = (\mathcal{P}_1, \mathcal{P}_0, \mathcal{U})^{\mathrm{IIb}}
= (\mathcal{U}, \mathcal{P}_0, \mathcal{P}_1)^{\mathrm{IIc}}$.

A fermion with spin $1/2$ in five dimensions is regarded as a Dirac fermion 
or a pair of Weyl fermions with opposite chiralities in four dimensions.
The representations of each Weyl fermions are decomposed in the same way,
but left-handed Weyl fermions and right-handed ones should have opposite $Z_2$ parities each other, i.e.,
$(\mathcal{P}_{0R},\mathcal{P}_{1R};\mathcal{U}_R)=(-\mathcal{P}_{0L},-\mathcal{P}_{1L};-\mathcal{U}_L)$,
from the requirement that the kinetic term is invariant under the $Z_2$ parity transformation.
Here, $(\mathcal{P}_{0R},\mathcal{P}_{1R};\mathcal{U}_R)$ and $(\mathcal{P}_{0L},\mathcal{P}_{1L};\mathcal{U}_L)$
are $Z_2$ parities for right-handed Weyl fermions and left-handed ones, respectively.
Zero modes for not only left-handed Weyl fermions but also right-handed ones, having even $Z_2$ parities, 
compose chiral fermions in the SM.

In SUSY models, the hypermultiplet is the fundamental quantity concerning bulk matter fields in five dimensions.
The hypermultiplet is equivalent to a pair of chiral multiplets 
with opposite gauge quantum numbers such as the representation $\mathbf{R}$ and 
the conjugate one $\overline{\mathbf{R}}$ in four dimensions.
The chiral multiplet with $\overline{\mathbf{R}}$ contains a left-handed Weyl fermion with $\overline{\mathbf{R}}_L$.
This Weyl fermion is regarded as a right-handed one with $\mathbf{R}_R$ by the use of the charge conjugation.
Hence our analysis works on SUSY models as well as non-SUSY ones.

\section{Unification of quarks and leptons based on $SO(2N)$}

Now let us investigate unification of quarks and leptons in $SO(2N)$ gauge theory on $S^1/Z_2$.
We count the numbers of fermion species coming from a single multiplet $\mathbf{2}_{1}^{N-1}$ or $\mathbf{2}_{2}^{N-1}$
based on the survival hypothesis for the following breaking patterns:
\begin{align}
& SO(2N) \to SO(10) \times H_1~,
\label{I-SO10}\\
& SO(2N) \to SO(6) \times SO(4) \times H_2 \simeq SU(4)_C\times SU(2)_L\times SU(2)_R \times H_2~,
\label{I-PS}\\
& SO(2N) \to SU(5) \times SU(N-5) \times U(1)^2~,
\label{II-SU5}
\end{align}
where $H_1$ and $H_2$ are some product groups such as $SO(2r_1) \times \cdots \times SO(2r_n)$.

\subsection{$SO(2N) \supset SO(10)$}

First, we study the symmetry breaking pattern $SO(2N) \to SO(10) \times H_1$.
In the case with the breaking pattern $SO(2N)\to SO(10)\times SO(2(N-5))$,
Weyl fermions with $\mathbf{2}_{1L}^{N-1}$ and $\mathbf{2}_{1R}^{N-1}$ are decomposed into
\begin{align}
\mathbf{2}_{1L}^{N-1}=(\mathbf{16},\mathbf{2}_1^{N-6})_L+(\overline{\mathbf{16}},\mathbf{2}_{2}^{N-6})_L~,~~ 
\mathbf{2}_{1R}^{N-1}=(\mathbf{16},\mathbf{2}_1^{N-6})_R+(\overline{\mathbf{16}},\mathbf{2}_{2}^{N-6})_R~,
\end{align}
and the $Z_2$ parities of each multiplet are given in Table \ref{SO(10)family1}.
\begin{table}[htbp]
\begin{center}
\begin{tabular}{c|c|c|c|c}
\hline
\multicolumn{2}{c|}{Representation}&$\mathcal{P}_0$&$\mathcal{P}_1$&$\mathcal{U}$\\ \hline \hline
$\mathbf{2}_{1L}^{N-1}$&$(\mathbf{16},\mathbf{2}_1^{N-6})_L$
&$+\eta_1^0$ &$+\eta_1^1$ &$+\eta_1^0\eta_1^1$ \\
&$(\overline{\mathbf{16}},\mathbf{2}_{2}^{N-6})_L$
&$+\eta_1^0$ &$-\eta_1^1$ &$-\eta_1^0\eta_1^1$ \\ \hline
$\mathbf{2}_{1R}^{N-1}$&$(\mathbf{16},\mathbf{2}_1^{N-6})_R$
&$-\eta_1^0$ &$-\eta_1^1$ &$+\eta_1^0\eta_1^1$ \\
&$(\overline{\mathbf{16}},\mathbf{2}_{2}^{N-6})_R$
&$-\eta_1^0$ &$+\eta_1^1$ &$-\eta_1^0\eta_1^1$ \\ \hline
\end{tabular}
\end{center}
\caption{$Z_2$ parity assignment for $\mathbf{2}_{1L}^{N-1}$ and $\mathbf{2}_{1R}^{N-1}$ in $SO(10) \times SO(2(N-5))$.}
\label{SO(10)family1}
\end{table}
If we take $\eta_1^1=\eta_1^0=+1$,
there appear no mirror particles and $(\mathbf{16},\mathbf{2}_1^{N-6})_L$ survives.
Then the number of $\mathbf{16}_L$ is regarded as that of families.
Hence we have $2^{N-6}$ families for the $SO(10)$ multiplets.
The same argument holds for the case with $\mathbf{2}_2^{N-1}$.

We find that no massless fermions survive in the case that $H_1 = SO(2r_1) \times SO(2r_2) \times SO(2r_3)$ or
$H_1 = SO(2r_1) \times SO(2r_2)$ after the survival hypothesis is imposed.

\subsection{$SO(2N)\supset SU(4)_C\times SU(2)_L\times SU(2)_R$}

Next, we study the symmetry breaking pattern $SO(2N) \to G_{\rm PS} \times H_2$
where $G_{\rm PS}$ is the Pati-Salam gauge group $SU(4)_C\times SU(2)_L\times SU(2)_R$~\cite{PS}.

In the case with $SO(2N)\to G_{\rm PS} \times SO(2q) \times SO(2s)$,
Weyl fermions with $\mathbf{2}_{1L}^{N-1}$ and $\mathbf{2}_{1R}^{N-1}$ are decomposed into
\begin{eqnarray}
&~&\mathbf{2}_{1L}^{N-1}
=(\mathbf{4},\mathbf{2},\mathbf{1},\mathbf{2}_1^{q-1},\mathbf{2}_1^{r-1})_L
+(\mathbf{4},\mathbf{2},\mathbf{1},\mathbf{2}_{2}^{q-1},\mathbf{2}_{2}^{r-1})_L \nonumber \\
&~&\hspace{30mm}+(\overline{\mathbf{4}},\mathbf{1},\mathbf{2},\mathbf{2}_1^{q-1},\mathbf{2}_1^{r-1})_L
+(\overline{\mathbf{4}},\mathbf{1},\mathbf{2},\mathbf{2}_{2}^{q-1},\mathbf{2}_{2}^{r-1})_L~, \nonumber \\
&~&~~~~~~~ +(\mathbf{4},\mathbf{1},\mathbf{2},\mathbf{2}_1^{q-1},\mathbf{2}_{2}^{r-1})_L
+(\mathbf{4},\mathbf{1},\mathbf{2},\mathbf{2}_{2}^{q-1},\mathbf{2}_1^{r-1})_L \notag \\
&~&\hspace{30mm}+(\overline{\mathbf{4}},\mathbf{2},\mathbf{1},\mathbf{2}_1^{q-1},\mathbf{2}_{2}^{r-1})_L
+(\overline{\mathbf{4}},\mathbf{2},\mathbf{1},\mathbf{2}_{2}^{q-1},\mathbf{2}_1^{r-1})_L~, \nonumber \\ 
&~&\mathbf{2}_{1R}^{N-1}
=(\mathbf{4},\mathbf{2},\mathbf{1},\mathbf{2}_1^{q-1},\mathbf{2}_1^{r-1})_R
+(\mathbf{4},\mathbf{2},\mathbf{1},\mathbf{2}_{2}^{q-1},\mathbf{2}_{2}^{r-1})_R \nonumber \\
&~&\hspace{30mm}+(\overline{\mathbf{4}},\mathbf{1},\mathbf{2},\mathbf{2}_1^{q-1},\mathbf{2}_1^{r-1})_R
+(\overline{\mathbf{4}},\mathbf{1},\mathbf{2},\mathbf{2}_{2}^{q-1},\mathbf{2}_{2}^{r-1})_R~, \nonumber \\
&~&~~~~~~~ +(\mathbf{4},\mathbf{1},\mathbf{2},\mathbf{2}_1^{q-1},\mathbf{2}_{2}^{r-1})_R
+(\mathbf{4},\mathbf{1},\mathbf{2},\mathbf{2}_{2}^{q-1},\mathbf{2}_1^{r-1})_R \nonumber \\
&~&\hspace{30mm}+(\overline{\mathbf{4}},\mathbf{2},\mathbf{1},\mathbf{2}_1^{q-1},\mathbf{2}_{2}^{r-1})_R
+(\overline{\mathbf{4}},\mathbf{2},\mathbf{1},\mathbf{2}_{2}^{q-1},\mathbf{2}_1^{r-1})_R~,
\end{eqnarray}
and the $Z_2$ parities of each multiplet are given in Table \ref{SU422-2}.
\begin{table}[htbp]
\begin{center}
\begin{tabular}{c|c|c|c|c}
\hline
\multicolumn{2}{c|}{Representation}&$\mathcal{P}_0$&$\mathcal{P}_1$&$\mathcal{U}$\\ \hline \hline
$\mathbf{2}_{1L}^{N-1}$
&$(\mathbf{4},\mathbf{2},\mathbf{1},\mathbf{2}_1^{r-1},\mathbf{2}_1^{s-1})_L$&
$+\eta_1^0$ &$+\eta_1^1$ &$+\eta_1^0\eta_1^1$ \\
&$(\mathbf{4},\mathbf{2},\mathbf{1},\mathbf{2}_{2}^{r-1},\mathbf{2}_{2}^{s-1})_L$&
$+\eta_1^0$ &$-\eta_1^1$ &$-\eta_1^0\eta_1^1$ \\
&$(\overline{\mathbf{4}},\mathbf{1},\mathbf{2},\mathbf{2}_1^{r-1},\mathbf{2}_1^{s-1})_L$&
$+\eta_1^0$ &$-\eta_1^1$ &$-\eta_1^0\eta_1^1$ \\
&$(\overline{\mathbf{4}},\mathbf{1},\mathbf{2},\mathbf{2}_{2}^{r-1},\mathbf{2}_{2}^{s-1})_L$&
$+\eta_1^0$ &$+\eta_1^1$ &$+\eta_1^0\eta_1^1$ \\
&$(\mathbf{4},\mathbf{1},\mathbf{2},\mathbf{2}_1^{r-1},\mathbf{2}_{2}^{s-1})_L$&
$-\eta_1^0$ &$+\eta_1^1$ &$-\eta_1^0\eta_1^1$ \\
&$(\mathbf{4},\mathbf{1},\mathbf{2},\mathbf{2}_{2}^{r-1},\mathbf{2}_1^{s-1})_L$&
$-\eta_1^0$ &$-\eta_1^1$ &$+\eta_1^0\eta_1^1$ \\
&$(\overline{\mathbf{4}},\mathbf{2},\mathbf{1},\mathbf{2}_1^{r-1},\mathbf{2}_{2}^{s-1})_L$&
$-\eta_1^0$ &$-\eta_1^1$ &$+\eta_1^0\eta_1^1$ \\
&$(\overline{\mathbf{4}},\mathbf{2},\mathbf{1},\mathbf{2}_{2}^{r-1},\mathbf{2}_1^{s-1})_L$&
$-\eta_1^0$ &$+\eta_1^1$ &$-\eta_1^0\eta_1^1$ \\ 
\hline 
$\mathbf{2}_{1R}^{N-1}$
&$(\mathbf{4},\mathbf{2},\mathbf{1},\mathbf{2}_1^{r-1},\mathbf{2}_1^{s-1})_R$&
$-\eta_1^0$ &$-\eta_1^1$ &$+\eta_1^0\eta_1^1$ \\
&$(\mathbf{4},\mathbf{2},\mathbf{1},\mathbf{2}_{2}^{r-1},\mathbf{2}_{2}^{s-1})_R$&
$-\eta_1^0$ &$+\eta_1^1$ &$-\eta_1^0\eta_1^1$ \\
&$(\overline{\mathbf{4}},\mathbf{1},\mathbf{2},\mathbf{2}_1^{r-1},\mathbf{2}_1^{s-1})_R$&
$-\eta_1^0$ &$+\eta_1^1$ &$-\eta_1^0\eta_1^1$ \\
&$(\overline{\mathbf{4}},\mathbf{1},\mathbf{2},\mathbf{2}_{2}^{r-1},\mathbf{2}_{2}^{s-1})_R$&
$-\eta_1^0$ &$-\eta_1^1$ &$+\eta_1^0\eta_1^1$ \\
&$(\mathbf{4},\mathbf{1},\mathbf{2},\mathbf{2}_1^{r-1},\mathbf{2}_{2}^{s-1})_R$&
$+\eta_1^0$ &$-\eta_1^1$ &$-\eta_1^0\eta_1^1$ \\
&$(\mathbf{4},\mathbf{1},\mathbf{2},\mathbf{2}_{2}^{r-1},\mathbf{2}_1^{s-1})_R$&
$+\eta_1^0$ &$+\eta_1^1$ &$+\eta_1^0\eta_1^1$ \\
&$(\overline{\mathbf{4}},\mathbf{2},\mathbf{1},\mathbf{2}_1^{r-1},\mathbf{2}_{2}^{s-1})_R$&
$+\eta_1^0$ &$+\eta_1^1$ &$+\eta_1^0\eta_1^1$ \\
&$(\overline{\mathbf{4}},\mathbf{2},\mathbf{1},\mathbf{2}_{2}^{r-1},\mathbf{2}_1^{s-1})_R$&
$+\eta_1^0$ &$-\eta_1^1$ &$-\eta_1^0\eta_1^1$ \\ 
\hline 
\end{tabular}
\end{center}
\caption{$Z_2$ parity assignment for $\mathbf{2}_{1L}^{N-1}$ and $\mathbf{2}_{1R}^{N-1}$ in 
$G_{\rm PS} \times SO(2q) \times SO(2s)$.}
\label{SU422-2}
\end{table}
If we take $\eta_1^1=\eta_1^0=+1$,
$(\mathbf{4},\mathbf{2},\mathbf{1},\mathbf{2}_1^{r-1},\mathbf{2}_1^{s-1})_L$,
$(\overline{\mathbf{4}},\mathbf{1},\mathbf{2},\mathbf{2}_{2}^{r-1},\mathbf{2}_{2}^{s-1})_L$,
$(\mathbf{4},\mathbf{1},\mathbf{2},\mathbf{2}_{2}^{r-1},\mathbf{2}_1^{s-1})_R$
and
$(\overline{\mathbf{4}},\mathbf{2},\mathbf{1},\mathbf{2}_1^{r-1},\mathbf{2}_{2}^{s-1})_R$
have zero modes.
Hence we have $2^{N-6}$ families for the $G_{\rm PS}$ multiplets.
The same argument holds for the case with $\mathbf{2}_2^{N-1}$.

In the case with $SO(2N)\to G_{\rm PS} \times SO(2(N-5))$,
Weyl fermions with $\mathbf{2}_{1L}^{N-1}$ and $\mathbf{2}_{1R}^{N-1}$ are decomposed into
\begin{align}
\mathbf{2}_{1L}^{N-1}&=(\mathbf{4},\mathbf{2},\mathbf{1},\mathbf{2}_1^{N-6})_L+(\overline{\mathbf{4}},\mathbf{1},\mathbf{2},\mathbf{2}_1^{N-6})_L
+(\mathbf{4},\mathbf{1},\mathbf{2},\mathbf{2}_{2}^{N-6})_L+(\overline{\mathbf{4}},\mathbf{2},\mathbf{1},\mathbf{2}_{2}^{N-6})_L~,\notag \\
\mathbf{2}_{1R}^{N-1}&=(\mathbf{4},\mathbf{2},\mathbf{1},\mathbf{2}_1^{N-6})_R+(\overline{\mathbf{4}},\mathbf{1},\mathbf{2},\mathbf{2}_1^{N-6})_R
+(\mathbf{4},\mathbf{1},\mathbf{2},\mathbf{2}_{2}^{N-6})_R+(\overline{\mathbf{4}},\mathbf{2},\mathbf{1},\mathbf{2}_{2}^{N-6})_R~,
\end{align}
and the $Z_2$ parities of each multiplet are given in Table \ref{SU422-1}.
\begin{table}[htbp]
\begin{center}
\begin{tabular}{c|c|c|c|c}
\hline
\multicolumn{2}{c|}{Representation}&$\mathcal{P}_0$&$\mathcal{P}_1$&$\mathcal{U}$\\ \hline \hline
$\mathbf{2}_{1L}^{N-1}$&$(\mathbf{4},\mathbf{2},\mathbf{1},\mathbf{2}_1^{N-6})_L$&
$+\eta_1^0$ &$+\eta_1^1$ &$+\eta_1^0\eta_1^1$ \\
&$(\overline{\mathbf{4}},\mathbf{1},\mathbf{2},\mathbf{2}_1^{N-6})_L$& 
$+\eta_1^0$ &$-\eta_1^1$ &$-\eta_1^0\eta_1^1$ \\
&$(\mathbf{4},\mathbf{1},\mathbf{2},\mathbf{2}_{2}^{N-6})_L$&
$-\eta_1^0$ &$-\eta_1^1$ &$+\eta_1^0\eta_1^1$ \\
&$(\overline{\mathbf{4}},\mathbf{2},\mathbf{1},\mathbf{2}_{2}^{N-6})_L$&
$-\eta_1^0$ &$+\eta_1^1$ &$-\eta_1^0\eta_1^1$ \\ 
\hline
$\mathbf{2}_{1R}^{N-1}$&$(\mathbf{4},\mathbf{2},\mathbf{1},\mathbf{2}_1^{N-6})_R$&
$-\eta_1^0$ &$-\eta_1^1$ &$+\eta_1^0\eta_1^1$ \\
&$(\overline{\mathbf{4}},\mathbf{1},\mathbf{2},\mathbf{2}_1^{N-6})_R$& 
$-\eta_1^0$ &$+\eta_1^1$ &$-\eta_1^0\eta_1^1$ \\
&$(\mathbf{4},\mathbf{1},\mathbf{2},\mathbf{2}_{2}^{N-6})_R$&
$+\eta_1^0$ &$+\eta_1^1$ &$+\eta_1^0\eta_1^1$ \\
&$(\overline{\mathbf{4}},\mathbf{2},\mathbf{1},\mathbf{2}_{2}^{N-6})_R$&
$+\eta_1^0$ &$-\eta_1^1$ &$-\eta_1^0\eta_1^1$ \\ \hline
\end{tabular}
\end{center}
\caption{$Z_2$ parity assignment for $\mathbf{2}_{1L}^{N-1}$ and $\mathbf{2}_{1R}^{N-1}$ in 
$G_{\rm PS} \times SO(2(N-5))$.}
\label{SU422-1}
\end{table}
If we take $\eta_1^1=\eta_1^0=+1$,
$(\mathbf{4},\mathbf{2},\mathbf{1},\mathbf{2}_1^{N-6})_L$ and $(\mathbf{4},\mathbf{1},\mathbf{2},\mathbf{2}_{2}^{N-6})_R$ have zero modes.
Then the number of such pairs is regarded as that of families.
Hence we have $2^{N-6}$ families for the $G_{\rm PS}$ multiplets.
The same argument holds for the case with $\mathbf{2}_2^{N-1}$.

\subsection{$SO(2N)\supset SU(5)$}

Finally, we study the symmetry breaking pattern $SO(2N) \to SU(5) \times SU(N-5) \times U(1)^2$.
In this case, Weyl fermions with $\mathbf{2}_{1L}^{N-1}$ and $\mathbf{2}_{1R}^{N-1}$ are decomposed into
\begin{align}
\mathbf{2}_{1L}^{N-1}
&=\sum_{k=0}^{[N/2]}\sum_{\ell=\mathrm{even}}({}_5C_{\ell},{}_{N-5}C_{2k-\ell})_L
+\sum_{k=0}^{[N/2]}\sum_{\ell=\mathrm{odd}}({}_5C_{\ell},{}_{N-5}C_{2k-\ell})_L~,\notag \\
\mathbf{2}_{1R}^{N-1}
&=\sum_{k=0}^{[N/2]}\sum_{\ell=\mathrm{even}}({}_5C_{\ell},{}_{N-5}C_{2k-\ell})_R
+\sum_{k=0}^{[N/2]}\sum_{\ell=\mathrm{odd}}({}_5C_{\ell},{}_{N-5}C_{2k-\ell})_R~,
\end{align}
and the $Z_2$ parities of each multiplet are given in Table \ref{SU(5)family1}.
\begin{table}[htbp]
\begin{center}
\begin{tabular}{c|c|c|c|c}
\hline
\multicolumn{2}{c|}{Representation}&$\mathcal{P}_0$&$\mathcal{P}_1$&$\mathcal{U}$\\ \hline \hline
$\mathbf{1}_L$&$({}_5C_0,{}_{N-5}C_{2k})_L$
&$+\eta_1^0$&$(-1)^k\eta_1^1$&$+(-1)^k\eta_1^0\eta_1^1$\\
$\mathbf{1}_R$&$({}_5C_0,{}_{N-5}C_{2k})_R$
&$-\eta_1^0$&$(-1)^k\eta_1^1$&$-(-1)^k\eta_1^0\eta_1^1$\\ \hline
$\mathbf{5}_L$&$({}_5C_1,{}_{N-5}C_{2k-1})_L$
&$-\eta_1^0$&$(-1)^k\eta_1^1$&$-(-1)^k\eta_1^0\eta_1^1$\\
$\mathbf{5}_R$&$({}_5C_1,{}_{N-5}C_{2k-1})_R$
&$+\eta_1^0$&$(-1)^k\eta_1^1$&$+(-1)^k\eta_1^0\eta_1^1$\\ \hline
$\mathbf{10}_L$&$({}_5C_2,{}_{N-5}C_{2k-2})_L$
&$+\eta_1^0$&$(-1)^k\eta_1^1$&$+(-1)^k\eta_1^0\eta_1^1$\\
$\mathbf{10}_R$&$({}_5C_2,{}_{N-5}C_{2k-2})_R$
&$-\eta_1^0$&$(-1)^k\eta_1^1$&$-(-1)^k\eta_1^0\eta_1^1$\\ \hline
$\overline{\mathbf{10}}_L$&$({}_5C_3,{}_{N-5}C_{2k-3})_L$
&$-\eta_1^0$&$(-1)^k\eta_1^1$&$-(-1)^k\eta_1^0\eta_1^1$\\
$\overline{\mathbf{10}}_R$&$({}_5C_3,{}_{N-5}C_{2k-3})_R$
&$+\eta_1^0$&$(-1)^k\eta_1^1$&$+(-1)^k\eta_1^0\eta_1^1$\\ \hline
$\overline{\mathbf{5}}_L$&$({}_5C_4,{}_{N-5}C_{2k-4})_L$
&$+\eta_1^0$&$(-1)^k\eta_1^1$&$+(-1)^k\eta_1^0\eta_1^1$\\
$\overline{\mathbf{5}}_R$&$({}_5C_4,{}_{N-5}C_{2k-4})_R$
&$-\eta_1^0$&$(-1)^k\eta_1^1$&$-(-1)^k\eta_1^0\eta_1^1$\\ \hline
$\overline{\mathbf{1}}_L$&$({}_5C_5,{}_{N-5}C_{2k-5})_L$
&$-\eta_1^0$&$(-1)^k\eta_1^1$&$-(-1)^k\eta_1^0\eta_1^1$\\
$\overline{\mathbf{1}}_R$&$({}_5C_5,{}_{N-5}C_{2k-5})_R$
&$+\eta_1^0$&$(-1)^k\eta_1^1$&$+(-1)^k\eta_1^0\eta_1^1$\\ \hline
\end{tabular}
\end{center}
\caption{$Z_2$ parity assignment for $\mathbf{2}_{1L}^{N-1}$ and $\mathbf{2}_{1R}^{N-1}$ in $SU(5) \times SU(N-5) \times U(1)^2$.}
\label{SU(5)family1}
\end{table}

Using the equivalence of $(\mathbf{5}_R)^c$ and $(\overline{\mathbf{10}}_R)^c$ 
with $\overline{\mathbf{5}}_L$ and $\mathbf{10}_L$, respectively,
the numbers of species $\mathbf{1}$, $\mathbf{10}_L$ and $\overline{\mathbf{5}}_L$ are given by
\begin{align}
&n_{\mathbf{1}}
=\sum_{k=\mathrm{even}}{}_{N-5}C_{2k}+\sum_{k=\mathrm{odd}}{}_{N-5}C_{2k-5}~,\\
&n_{\mathbf{10}_L}
=\sum_{k=\mathrm{even}}{}_{N-5}C_{2k-2}+\sum_{k=\mathrm{odd}}{}_{N-5}C_{2k-3}~,\\
&n_{\overline{\mathbf{5}}_L}
=\sum_{k=\mathrm{even}}{}_{N-5}C_{2k-4}+\sum_{k=\mathrm{odd}}{}_{N-5}C_{2k-1}~, 
\end{align}
in the case with $\eta_1^0=\eta_1^1=+1$ and 
\begin{align}
&n_{\mathbf{1}}
=\sum_{k=\mathrm{odd}}{}_{N-5}C_{2k}+\sum_{k=\mathrm{even}}{}_{N-5}C_{2k-5}~, \\
&n_{\mathbf{10}_L}
=\sum_{k=\mathrm{odd}}{}_{N-5}C_{2k-2}+\sum_{k=\mathrm{even}}{}_{N-5}C_{2k-3}~, \\
&n_{\overline{\mathbf{5}}_L}
=\sum_{k=\mathrm{odd}}{}_{N-5}C_{2k-4}+\sum_{k=\mathrm{even}}{}_{N-5}C_{2k-1}~,
\end{align}
in the case with $\eta_1^0=-\eta_1^1=+1$.
Here $n_{\mathbf{1}}$ is the total number of $SU(5)$ singlets $\mathbf{1}$.
They are regarded as the so-called right-handed neutrinos 
which can obtain heavy Majorana masses among themselves as well as the Dirac masses with left-handed neutrinos.
Some of them can be involved in see-saw mechanism~\cite{see-saw}.

In the same way, Weyl fermions with $\mathbf{2}_{2L}^{N-1}$ and $\mathbf{2}_{2R}^{N-1}$ are decomposed into
\begin{align}
\mathbf{2}_{2L}^{N-1}
&=\sum_{k=0}^{[(N-1)/2]}\sum_{\ell=\mathrm{even}}({}_5C_{\ell},{}_{N-5}C_{2k-\ell+1})_L
+\sum_{k=0}^{[(N-1)/2]}\sum_{\ell=\mathrm{odd}}({}_5C_{\ell},{}_{N-5}C_{2k-\ell+1})_L~,\notag \\
\mathbf{2}_{2R}^{N-1}
&=\sum_{k=0}^{[(N-1)/2]}\sum_{\ell=\mathrm{even}}({}_5C_{\ell},{}_{N-5}C_{2k-\ell+1})_R
+\sum_{k=0}^{[(N-1)/2]}\sum_{\ell=\mathrm{odd}}({}_5C_{\ell+1},{}_{N-5}C_{2k-\ell})_R~,
\end{align}
and the $Z_2$ parities of each multiplet are in Table \ref{SU(5)family2}.
\begin{table}[htbp]
\begin{center}
\begin{tabular}{c|c|c|c|c}
\hline
\multicolumn{2}{c|}{Representation}&$\mathcal{P}_0$&$\mathcal{P}_1$&$\mathcal{U}$\\ \hline \hline
$\mathbf{1}_L$&$({}_5C_0,{}_{N-5}C_{2k+1})_L$
&$+\eta_2^0$&$(-1)^k\eta_2^1$&$+(-1)^k\eta_2^0\eta_2^1$\\
$\mathbf{1}_R$&$({}_5C_0,{}_{N-5}C_{2k+1})_R$
&$-\eta_2^0$&$(-1)^k\eta_2^1$&$-(-1)^k\eta_2^0\eta_2^1$\\ \hline
$\mathbf{5}_L$&$({}_5C_1,{}_{N-5}C_{2k})_L$
&$-\eta_2^0$&$(-1)^k\eta_2^1$&$-(-1)^k\eta_2^0\eta_2^1$\\
$\mathbf{5}_R$&$({}_5C_1,{}_{N-5}C_{2k})_R$
&$+\eta_2^0$&$(-1)^k\eta_2^1$&$+(-1)^k\eta_2^0\eta_2^1$\\ \hline
$\mathbf{10}_L$&$({}_5C_2,{}_{N-5}C_{2k-1})_L$
&$+\eta_2^0$&$(-1)^k\eta_2^1$&$+(-1)^k\eta_2^0\eta_2^1$\\
$\mathbf{10}_R$&$({}_5C_2,{}_{N-5}C_{2k-1})_R$
&$-\eta_2^0$&$(-1)^k\eta_2^1$&$-(-1)^k\eta_2^0\eta_2^1$\\ \hline
$\overline{\mathbf{10}}_L$&$({}_5C_3,{}_{N-5}C_{2k-2})_L$
&$-\eta_2^0$&$(-1)^k\eta_2^1$&$-(-1)^k\eta_2^0\eta_2^1$\\
$\overline{\mathbf{10}}_R$&$({}_5C_3,{}_{N-5}C_{2k-2})_R$
&$+\eta_2^0$&$(-1)^k\eta_2^1$&$+(-1)^k\eta_2^0\eta_2^1$\\ \hline
$\overline{\mathbf{5}}_L$&$({}_5C_4,{}_{N-5}C_{2k-3})_L$
&$+\eta_2^0$&$(-1)^k\eta_2^1$&$+(-1)^k\eta_2^0\eta_2^1$\\
$\overline{\mathbf{5}}_R$&$({}_5C_4,{}_{N-5}C_{2k-3})_R$
&$-\eta_2^0$&$(-1)^k\eta_2^1$&$-(-1)^k\eta_2^0\eta_2^1$\\ \hline
$\overline{\mathbf{1}}_L$&$({}_5C_5,{}_{N-5}C_{2k-4})_L$
&$-\eta_2^0$&$(-1)^k\eta_2^1$&$-(-1)^k\eta_2^0\eta_2^1$\\
$\overline{\mathbf{1}}_R$&$({}_5C_5,{}_{N-5}C_{2k-4})_R$
&$+\eta_2^0$&$(-1)^k\eta_2^1$&$+(-1)^k\eta_2^0\eta_2^1$\\ \hline
\end{tabular}
\end{center}
\caption{$Z_2$ parity assignment for $\mathbf{2}_{2L}^{N-1}$ and $\mathbf{2}_{2R}^{N-1}$ in $SU(5) \times SU(N-5) \times U(1)^2$.}
\label{SU(5)family2}
\end{table}
The numbers of species $\mathbf{1}$, $\mathbf{10}_L$ and $\overline{\mathbf{5}}_L$ are given by
\begin{align}
&n_{\mathbf{1}} 
=\sum_{k=\mathrm{even}}{}_{N-5}C_{2k+1}+\sum_{k=\mathrm{odd}}{}_{N-5}C_{2k-4}~, \\
&n_{\mathbf{10}_L}
=\sum_{k=\mathrm{even}}{}_{N-5}C_{2k-1}+\sum_{k=\mathrm{odd}}{}_{N-5}C_{2k-2}~, \\
&n_{\overline{\mathbf{5}}_L}
=\sum_{k=\mathrm{even}}{}_{N-5}C_{2k-3}+\sum_{k=\mathrm{odd}}{}_{N-5}C_{2k}~,
\end{align}
in the case with $\eta_2^0=\eta_2^1=+1$ and
\begin{align}
&n_{\mathbf{1}} 
=\sum_{k=\mathrm{odd}}{}_{N-5}C_{2k+1}+\sum_{k=\mathrm{even}}{}_{N-5}C_{2k-4}~, \\
&n_{\mathbf{10}_L} 
=\sum_{k=\mathrm{odd}}{}_{N-5}C_{2k-1}+\sum_{k=\mathrm{even}}{}_{N-5}C_{2k-2}~, \\
&n_{\overline{\mathbf{5}}_L}
=\sum_{k=\mathrm{odd}}{}_{N-5}C_{2k-3}+\sum_{k=\mathrm{even}}{}_{N-5}C_{2k}~,
\end{align}
in the case with $\eta_2^0=-\eta_2^1=+1$.

As examples, the numbers of species $\mathbf{1}$, $\mathbf{10}_L$ and $\overline{\mathbf{5}}_L$ 
for $SO(14)$, $SO(16)$ and $SO(18)$ are listed in 
Tables \ref{SO(14)family}, \ref{SO(16)family} and \ref{SO(18)family}.
\begin{table}[htbp]
\begin{center}
\begin{tabular}{c|c|c|c|c}
\hline
&$\eta_1^0=\eta_1^1=+1$&$\eta_1^0=-\eta_1^1=+1$&$\eta_2^0=\eta_2^1=+1$&$\eta_2^0=-\eta_2^1=+1$\\ \hline \hline
$n_{\mathbf{1}}$&3&1&3&1\\ \hline
$n_{\mathbf{10}_L}$&1&3&1&3\\ \hline
$n_{\overline{\mathbf{5}}_L}$&3&1&3&1\\ \hline
\end{tabular}
\end{center}
\caption{The numbers of $\mathbf{1}$, $\mathbf{10}_L$ and $\overline{\mathbf{5}}_L$ for $SO(14)$.}
\label{SO(14)family}
\end{table}
\begin{table}[htbp]
\begin{center}
\begin{tabular}{c|c|c|c|c}
\hline
& $\eta_1^0=\eta_1^1=+1$&$\eta_1^0=-\eta_1^1=+1$&$\eta_2^0=\eta_2^1=+1$&$\eta_2^0=-\eta_2^1=+1$\\ \hline \hline
$n_{\mathbf{1}}$&4&4&6&2\\ \hline
$n_{\mathbf{10}_L}$&4&4&2&6\\ \hline
$n_{\overline{\mathbf{5}}_L}$&4&4&6&2\\ \hline
\end{tabular}
\end{center}
\caption{The numbers of $\mathbf{1}$, $\mathbf{10}_L$ and $\overline{\mathbf{5}}_L$ for $SO(16)$.}
\label{SO(16)family}
\end{table}
\begin{table}[htbp]
\begin{center}
\begin{tabular}{c|c|c|c|c}
\hline
& $\eta_1^0=\eta_1^1=+1$&$\eta_1^0=-\eta_1^1=+1$&$\eta_2^0=\eta_2^1=+1$&$\eta_2^0=-\eta_2^1=+1$\\ \hline \hline
$n_{\mathbf{1}}$&6&10&10&6\\ \hline
$n_{\mathbf{10}_L}$&10&6&6&10\\ \hline
$n_{\overline{\mathbf{5}}_L}$&6&10&10&6\\ \hline
\end{tabular}
\end{center}
\caption{The numbers of $\mathbf{1}$, $\mathbf{10}_L$ and $\overline{\mathbf{5}}_L$ for $SO(18)$.}
\label{SO(18)family}
\end{table}

\section{Conclusions and discussions}

We have studied the possibility of family unification on the basis of $SO(2N)$ gauge theory
on the five-dimensional space-time, $M^4\times S^1/Z_2$.
We have found that several $SO(10)$, $SU(4) \times SU(2)_L \times SU(2)_R$ or $SU(5)$ multiplets
come from a single bulk multiplet of $SO(2N)$ after the orbifold breaking
and obtained the numbers of species.
Other multiplets including brane fields are necessary to compose
three families of quarks and leptons.
Our results can give a starting point for the construction of a more realistic model.

There are several open questions, which are left for future work.

The unwanted matter degrees of freedom can be successfully made massive thanks to the orbifolding. 
However, some extra gauge fields remain massless even after the symmetry breaking due to the Hosotani mechanism. 
In most cases, this kind of non-abelian gauge subgroup plays the role of family symmetry.
These massless degrees of freedom must be made massive by further breaking of the family symmetry.
Here, we point out that brane fields can be key to the solutions.
Most models have chiral anomalies at the four-dimensional boundaries and
we have a choice to introduce appropriate brane fields to cancel these anomalies.
Further, scalar components of some brane superfields can play a role of Higgs fields for the breakdown of extra gauge symmetries 
including non-abelian gauge symmetries.
As a result, extra massless fields including the family gauge bosons can be massive.

In general, there appear $D$-term contributions to scalar masses in supersymmetric models 
after the breakdown of such extra gauge symmetries and the $D$-term contributions lift the mass degeneracy.~\cite{D,H&K,KM&Y}.
The mass degeneracy for each squark and slepton species in the first two families is favorable for suppressing
flavor-changing neutral current (FCNC) processes.
The dangerous FCNC processes can be avoided if the sfermion masses in the first two families are rather large 
or the fermion and its superpartner mass matrices are aligned.
The requirement of degenerate masses would yield a constraint on the $D$-term condensations
and/or SUSY breaking mechanism unless other mechanisms work.
If we consider the Scherk-Schwarz mechanism~\cite{S&S} for $N=1$ SUSY breaking, the $D$-term condensations
can vanish for the gauge symmetries broken at the orbifold breaking scale,
because of a universal structure of the soft SUSY breaking parameters.
The $D$-term contributions have been studied in the framework of $SU(N)$ orbifold GUTs~\cite{KK}.


Fermion mass hierarchy and generation mixings can also occur through the Froggatt-Nielsen mechanism~\cite{FN}
on the breakdown of extra gauge symmetries and the suppression of brane-localized Yukawa coupling constants 
among brane weak Higgs doublets and bulk matters with the volume suppression factor~\cite{Y}.

The orbifold GUT is more naturally realized in warped space, see e.g.~\cite{Nomura:2004zs} for a review. 
The Hosotani mechanism has been studied in warped space~\cite{warpDSB}
and it has been applied on the gauge-Higgs unification~\cite{GaugeHiggs}.
It would be interesting to look for the orbifold family unification based on warped space
and/or other types of orbifolds.\footnote{
Equivalence classes of BCs in $SU(N)$ gauge theory have been studied based on six-dimensional space-time including
$T^2/Z_2$ in Ref.~\cite{HNT} and other two-dimensional orbifolds in Ref.~\cite{KM}.}

It would be interesting to study cosmological implications of the class of models presented in this paper, 
see e.g.~\cite{Khlopov:1999rs} and references therein for useful articles toward this direction.

\section*{Acknowledgements}
This work was supported in part by scientific grants from the Ministry of Education, Culture, Sports, Science and Technology
under Grant Nos.18204024 and 18540259 (Y.K.) .

\appendix

\section{Gauge invariance and equivalence relations}

We discuss the gauge invariance on $S^1/Z_2$. 
Given the BCs $(P_0, P_1, U)$, there still remains residual gauge invariance.
Under the gauge transformation with $\Omega(x,y)$, $A_M$ transforms as
\begin{align}
A_M \to A'_M = \Omega A_M \Omega^{-1} - \frac{i}{g}\Omega \partial_M \Omega^{-1}~,
\label{gauge-tr}
\end{align}
where $g$ is a gauge coupling and $A'_M$ satisfies, instead of  (\ref{gauge-BCs}),
\begin{align}
& s_0:~A'_{\mu}(x,-y)=P'_0A'_{\mu}(x,y){P'}_0^{-1}~,~~ A'_{y}(x,-y)=-P'_0A'_{y}(x,y){P'}_0^{-1}~,  \notag \\
& s_1:~A'_{\mu}(x,2\pi R-y)=P'_1A'_{\mu}(x,y){P'}_1^{-1}~,~~ A'_{y}(x,2\pi R-y)=-P'_1A'_{y}(x,y){P'}_1^{-1}~, \notag \\ 
& t:~A'_{M}(x, y+2\pi R)=U' A'_{M}(x,y) {U'}^{-1}~.
\label{gauge-BCs-prime}
\end{align}
The $P'_{0}$, $P'_{1}$ and $U'$ are given by 
\begin{align}
&P'_{0}=\Omega(x, -y)P_{0}\Omega^{-1}(x, y)~, ~~
P'_{1}=\Omega(x, 2\pi R-y)P_{1}\Omega^{-1}(x, y)~,\notag \\
&U'=\Omega(x, y+2\pi R)U\Omega^{-1}(x, y)~,
\end{align} 
where we assume that $P'_{0}$, $P'_{1}$ and $U'$ are independent of $y$.

Theories with different BCs should be equivalent with regard to physical content if they are connected by 
gauge transformations.
The key observation is that the physics should not depend on the gauge chosen.
The equivalence is guaranteed in the Hosotani mechanism\cite{H}
and the two sets of BCs are equivalent:
\begin{align}
(P_{0}, P_{1}, U) \sim (P'_{0}, P'_1, U')~.
\label{Eq-rel}
\end{align}
The equivalence relation (\ref{Eq-rel}) defines equivalence classes of the BCs.

The physical symmetry is understood from the analysis including the Wilson line phases as follows.
The Wilson line phases are phases of $WU$ given by
\begin{align}
WU = \mathbf{P}\exp\left(ig\int_C A_y(x,y)dy\right)\cdot U~,
\label{WU}
\end{align}
where $\mathbf{P}$ is path-ordering along a non-contractible loop on $S^1$.
The eigenvalues of $WU$ are gauge invariant and become physical degrees of freedom.
The dynamical phases are given by $\theta^b = 2 \pi R g A_y^b$ related to
the generators $T^b$ which anticommute with $(P_0, P_1)$, i.e., $\{T^b, P_0\} = \{T^b, P_1\} =0$.
They correspond to the parts of $A_y$ with even $Z_2$ parities.
The physical vacuum is given by the configuration of $\theta^b$ which minimizes the effective potential.
Suppose that the effective potential is minimized at $\langle  A_y \rangle$ 
such that $W \equiv \exp(ig 2\pi R  \langle A_y \rangle) \neq I$ with $(P_0, P_1, U)$.
Perform the gauge transformation with
$\Omega = \exp[i g (y + \alpha) \langle A_y \rangle]$,
which brings $\langle A_y \rangle$ to $\langle {A'}_y \rangle = 0$.
Then the BCs change to
\begin{eqnarray}
\hspace{-0.3cm} (P_0^{\rm sym}, P_1^{\rm sym}, U^{\rm sym})
\equiv (P'_0, P'_1, U') = (e^{2ig\alpha\langle A_y \rangle} P_0,
e^{2ig(\alpha + \pi R)\langle A_y \rangle} P_1, e^{ig 2\pi R \langle A_y \rangle} U (=WU))~.
\label{equiv3}
\end{eqnarray}
Since $\langle {A'}_y \rangle$ vanishes in the new gauge, 
the unbroken symmetry is spanned by the generators $T^a$
which commute with $(P_0^{\rm sym}, P_1^{\rm sym})$, i.e., $[T^a, P_0^{\rm sym}] = [T^a, P_1^{\rm sym}] =0$.

Let us derive the equivalence relations among BCs based on Type-I and Type-II.
We consider $SO(4)$ gauge theory.
For the gauge transformation with $\Omega(y)$ given by
\begin{align}
\Omega(y) =\exp\left[i(a\sigma_1\otimes \tau_2 +b\sigma_3 \otimes \tau_2)y/2\pi R\right]~,
\label{Omega1}
\end{align}
we find the equivalence relations:
\begin{align}
&\mbox{Type-I}:(\sigma_0 \otimes \tau_3, \sigma_0 \otimes \tau_3)
\sim (\sigma_0 \otimes \tau_3, \exp\left[i(a\sigma_1 \otimes \tau_2 +b\sigma_3 \otimes \tau_2 )\right]\sigma_0 \otimes \tau_3)~,\notag \\
&\mbox{Type-IIa}:(\sigma_0 \otimes \tau_3, \sigma_2 \otimes \tau_0)
\sim (\sigma_0 \otimes \tau_3, \exp\left[i(a\sigma_1 \otimes \tau_2 +b\sigma_3 \otimes \tau_2 )\right]\sigma_2 \otimes \tau_0)~,\notag \\
&\mbox{Type-IIb}:(\pm \sigma_2 \otimes \tau_0, \sigma_0 \otimes \tau_3)
\sim (\pm \sigma_2 \otimes \tau_0, \exp\left[i(a\sigma_1 \otimes \tau_2 +b\sigma_3 \otimes \tau_2 )\right]\sigma_0 \otimes \tau_3)~,\notag \\
&\mbox{Type-IIc}:(\pm \sigma_2 \otimes \tau_0, \sigma_2 \otimes \tau_0)
\sim (\pm \sigma_2 \otimes \tau_0, \exp\left[i(a\sigma_1 \otimes \tau_2 +b\sigma_3 \otimes \tau_2 )\right]\sigma_2 \otimes \tau_0)~,
\end{align}
where $a,b \in \mathbb{R}$ and $\tau_i$ $(i=1,2,3)$ are also Pauli matrices.
When $\sqrt{a^2+b^2}=\pi \mod{2\pi}$, the equivalence relations become as
\begin{align}
&\mbox{Type-I}:(\sigma_0 \otimes \tau_3, \sigma_0 \otimes \tau_3)
\sim (\sigma_0 \otimes \tau_3, -\sigma_0 \otimes \tau_3)~,\notag \\
&\mbox{Type-IIa}:(\sigma_0 \otimes \tau_3, \sigma_2 \otimes \tau_0)
\sim (\sigma_0 \otimes \tau_3, -\sigma_2 \otimes \tau_0)~,\notag \\
&\mbox{Type-IIb}:(\pm \sigma_2 \otimes \tau_0, \sigma_0 \otimes \tau_3)
\sim (\pm \sigma_2 \otimes \tau_0, -\sigma_0 \otimes \tau_3)~,\notag \\
&\mbox{Type-IIc}:(\pm \sigma_2 \otimes \tau_0, \sigma_2 \otimes \tau_0)
\sim (\pm \sigma_2 \otimes \tau_0, -\sigma_2 \otimes \tau_0)~.
\label{EQ1}
\end{align}
Because $(\pm \sigma_2 \otimes \tau_0, -\sigma_0 \otimes \tau_3)$ equals to $(\pm \sigma_2 \otimes \tau_0, \sigma_0 \otimes \tau_3)$,
we obtain no relation concerning to (\ref{Omega1}) for Type-IIb.
Using (\ref{EQ1}), the following relations in $SO(2N)$ gauge theory are derived,
\begin{align}
[p,q;r,s]^{\mathrm{I}}~
&\sim [p-1,q+1;r+1,s-1]^{\mathrm{I}}~~~\mathrm{for}~p,s\geq 1~,\notag \\
&\sim [p+1,q-1;r-1,s+1]^{\mathrm{I}}~~~\mathrm{for}~q,r\geq 1~,\\
[p,q;r,s]^{\mathrm{IIa}} 
&\sim [p-1,q+1;r-1,s+1]^{\mathrm{IIa}}~~~\mathrm{for}~p,s\geq 1~,\notag \\
&\sim [p+1,q-1;r+1,s-1]^{\mathrm{IIa}}~~~\mathrm{for}~q,r\geq 1~,\\
[p,q;r,s]^{\mathrm{IIc}} 
&\sim [p-2,q+2;r,s]^{\mathrm{IIc}}~~~\mathrm{for}~p\geq 2~,\notag \\
&\sim [p+2,q-2;r,s]^{\mathrm{IIc}}~~~\mathrm{for}~q\geq 2~,\notag \\
&\sim [p,q;r-2,s+2]^{\mathrm{IIc}}~~~\mathrm{for}~r\geq 2~,\notag \\
&\sim [p,q;r+2,s-2]^{\mathrm{IIc}}~~~\mathrm{for}~s\geq 2~.
\end{align}
For another gauge transformation with $\Omega(y)$ given by
\begin{align}
\Omega(y) =\exp\left[i(a\sigma_2 \otimes \tau_1 +b\sigma_0 \otimes \tau_2 )y/2\pi R\right]~,
\end{align}
we find the equivalence relations:
\begin{align}
&\mbox{Type-I}:(\sigma_0 \otimes \tau_3, \sigma_0 \otimes \tau_3)
\sim (\sigma_0 \otimes \tau_3, \exp\left[i(a\sigma_2 \otimes \tau_1 +b\sigma_0 \otimes \tau_2 )\right]\sigma_0 \otimes \tau_3)~,\notag \\
&\mbox{Type-IIa}:(\sigma_0 \otimes \tau_3, \sigma_2 \otimes \tau_3)
\sim (\sigma_0 \otimes \tau_3, \exp\left[i(a\sigma_2 \otimes \tau_1 +b\sigma_0 \otimes \tau_2 )\right]\sigma_2 \otimes \tau_3)~,\notag \\
&\mbox{Type-IIb}:(\sigma_2 \otimes \tau_3, \sigma_0 \otimes \tau_3)
\sim (\sigma_2 \otimes \tau_3, \exp\left[i(a\sigma_2 \otimes \tau_1 +b\sigma_0 \otimes \tau_2 )\right]\sigma_0 \otimes \tau_3)~,\notag \\
&\mbox{Type-IIc}:(\sigma_2 \otimes \tau_3, \sigma_2 \otimes \tau_3)
\sim (\sigma_2 \otimes \tau_3, \exp\left[i(a\sigma_2 \otimes \tau_1 +b\sigma_0 \otimes \tau_2 )\right]\sigma_2 \otimes \tau_3)~.
\end{align}
When $\sqrt{a^2+b^2}=\pi \mod{2\pi}$, the equivalence relations become as
\begin{align}
&\mbox{Type-I}:(\sigma_0 \otimes \tau_3, \sigma_0 \otimes \tau_3)
\sim (\sigma_0 \otimes \tau_3, -\sigma_0 \otimes \tau_3)~,\notag \\
&\mbox{Type-IIa}:(\sigma_0 \otimes \tau_3, \sigma_2 \otimes \tau_3)
\sim (\sigma_0 \otimes \tau_3, -\sigma_2 \otimes \tau_3)~,\notag \\
&\mbox{Type-IIb}:(\sigma_2 \otimes \tau_3, \sigma_0 \otimes \tau_3)
\sim (\sigma_2 \otimes \tau_3, -\sigma_0 \otimes \tau_3)~,\notag \\
&\mbox{Type-IIc}:(\sigma_2 \otimes \tau_3, \sigma_2 \otimes \tau_3)
\sim (\sigma_2 \otimes \tau_3, -\sigma_2 \otimes \tau_3)~.
\label{EQ2}
\end{align}
Using (\ref{EQ2}), the following relations in $SO(2N)$ gauge theory are derived,
\begin{align}
[p,q;r,s]^{\mathrm{I}}~
&\sim [p-1,q+1;r+1,s-1]^{\mathrm{I}}~~~\mathrm{for}~p,s\geq 1~,\notag \\
&\sim [p+1,q-1;r-1,s+1]^{\mathrm{I}}~~~\mathrm{for}~q,r\geq 1~,\\
[p,q;r,s]^{\mathrm{IIa}} 
&\sim [p-1,q+1;r+1,s-1]^{\mathrm{IIa}}~~~\mathrm{for}~p,s\geq 1~,\notag \\
&\sim [p+1,q-1;r-1,s+1]^{\mathrm{IIa}}~~~\mathrm{for}~q,r\geq 1~,\\
[p,q;r,s]^{\mathrm{IIb}}
&\sim [p-1,q+1;r+1,s-1]^{\mathrm{IIb}}~~~\mathrm{for}~p,s\geq 1~,\notag \\
&\sim [p+1,q-1;r-1,s+1]^{\mathrm{IIb}}~~~\mathrm{for}~q,r\geq 1~,\\
[p,q;r,s]^{\mathrm{IIc}} 
&\sim [p-1,q+1;r+1,s-1]^{\mathrm{IIc}}~~~\mathrm{for}~p,s\geq 1~,\notag \\
&\sim [p+1,q-1;r-1,s+1]^{\mathrm{IIc}}~~~\mathrm{for}~q,r\geq 1~.
\end{align}

\section{$S^1/Z_2$ Orbifold breaking of $SO(2N+1)$}

We study the orbifold symmetry breaking in $SO(2N+1)$.
Because $SO(2N+1)\supset SO(2N)$,  
the generators of $SO(2N+1)$ are written as
\begin{align}
\left(
\begin{array}{c|c}
{\text{\Large{e{\sl so}(2{\sl N})f}}}& (*) \\ \hline
(*)^t & 0 
\end{array}
\right)~,
\end{align}
where e$so(2N)$f represents generators of $SO(2N)$ and $(*)$ are $2N \times 1$ matrix.

As an example, let us take the following representation matrices:
\begin{align}
P_0 =
\left(
\begin{array}{c|c}
{\sigma_0 \otimes I_N}&0\\ \hline
0& -1
\end{array}
\right)~,~~ 
P_1 =
\left(
\begin{array}{c|c}
{\sigma_0 \otimes I_{m,n}}&0\\ \hline
0& \eta
\end{array}
\right)~,
\end{align}
where $\eta=\pm 1$.
Then we obtain the breaking pattern:
\begin{align}
SO(2N+1)\rightarrow SO(2m)\times SO(2n)~,
\end{align}
and the $Z_2$ parities for gauge bosons $A_{\mu}^{\alpha}$ are assigned as
\begin{align}
\mathbf{N(2N+1)} &= (\mathbf{m(2m-1)},\mathbf{1})^{++;+}
+(\mathbf{1},\mathbf{n(2n-1)})^{++;+}
+(\mathbf{2m},\mathbf{2n})^{+-;-} \notag\\
&~~ +(\mathbf{2m},\mathbf{1})^{-\mp;\pm}
+(\mathbf{1},\mathbf{2n})^{-\pm;\mp}~.
\label{SO2N+1Gparity}
\end{align}
There is one spinor representation ${\mathbf{2}}^{N}$ in $SO(2N+1)$, which is decomposed into
\begin{align}
&\mathbf{2}_L^{N}=(\mathbf{2}_{1}^{m-1},\mathbf{2}_{1}^{n-1})_L^{++;+} + (\mathbf{2}_{2}^{m-1},\mathbf{2}_{2}^{n-1})_L^{+-;-}
   + (\mathbf{2}_{1}^{m-1},\mathbf{2}_{2}^{n-1})_L^{-\pm;\mp} + (\mathbf{2}_{2}^{m-1},\mathbf{2}_{1}^{n-1})_L^{-\mp;\pm}~, \\
&\mathbf{2}_R^{N}=(\mathbf{2}_{1}^{m-1},\mathbf{2}_{1}^{n-1})_R^{--;-} + (\mathbf{2}_{2}^{m-1},\mathbf{2}_{2}^{n-1})_R^{-+;+}
   + (\mathbf{2}_{1}^{m-1},\mathbf{2}_{2}^{n-1})_R^{+\mp;\pm} + (\mathbf{2}_{2}^{m-1},\mathbf{2}_{1}^{n-1})_R^{+\pm;\mp}~,
\end{align}
where we take an appropriate intrinsic $Z_2$ parity assignment.
Using the above assignment, we find $2^{N-5}$ families with $\eta = +1$ and no family with $\eta=-1$ 
for $SO(10)$ multiplets $\mathbf{16}_L$
after the breaking $SO(2N+1)\rightarrow SO(10)\times SO(2(N-5))$.


\end{document}